\documentclass[letterpaper]{aipproc}

\usepackage{subfigure}
\usepackage{mathptmx}
\usepackage{latexsym}
\makeatletter
\newcommand\setcaptype[1]{\def\@captype{#1}}
\makeatother

\layoutstyle{6x9} 

\SetInternalRegister\hbadness{8000} 

\newcommand\doingARLO[2][]{%
  \ifx\mmref\undefined #1\else #2\fi
}
\newcommand\ion[2]{#1$\;${\small\rmfamily\@{#2}}\relax}%
\def\gtrsim{\mathrel{\hbox{\rlap{\hbox{\lower4pt\hbox{$\sim$}}}\hbox{$>$}}}}

\begin{document}

\title 
      [Astrobiology]
      {Astrobiology:  An Astronomer's Perspective}


\author{Edwin A. Bergin}{
  address={University of Michigan, Department of Astronomy, 500 Church Street, Ann Arbor MI 48109, USA},
  email={ebergin@umich.edu},
}

\copyrightyear  {2013}

\begin{abstract}
In this review we explore  aspects of the field of astrobiology from an astronomical viewpoint.
We therefore focus on the origin of life in the context of planetary formation, with  additional emphasis on
tracing the most abundant volatile elements, C, H, O,  and N that are used by life on Earth.   We first explore
the history of life on our planet and outline the current state of our knowledge regarding the delivery
of the C, H, O, N elements to the Earth.   We then discuss how astronomers track the gaseous and solid
molecular carriers of these volatiles throughout the process of star and planet formation.
It is now clear that the early stages of star formation fosters the
creation of water and simple organic molecules with enrichments of heavy isotopes.   
These molecules are found as ice coatings on the solid materials 
that represent microscopic beginnings  of terrestrial worlds.  Based on the meteoritic and cometary record, the process of planet formation, and the local environment, lead to additional increases in organic complexity.  The astronomical connections towards this stage are only now being directly made.    Although the exact details are uncertain, it is likely that the birth process of star and planets  
likely leads to terrestrial worlds being born with abundant water and organics on the surface. 
\end{abstract}

\date{\today}

\maketitle

\section{Introduction}

The field of astrobiology seeks to explore the question of our origin and, more broadly,
the question of life in the universe.   
This exploration must start with the
birth of water-coated terrestrial worlds that had carbon present in some organic form.   
Within an uncertain and likely variable micro- and macro-environment, the genesis of biotic or reproductive chemistry arises from abiotic chemical reactions.
From such biotic precursors ensues an evolutionary march towards complexity, which, over the course of billions of years, yields the life-rich planet we know today.
As one might imagine, this presents a vast academic landscape
and a single publication will  not capture the multitude of  associated questions.   

From the perspective of an 
 astronomer we can ask what elements of ``astro-'' contribute to astrobiology.  One key astronomically motivated question is how common is life in our galaxy?  Since life as we know it requires liquid water as a medium for biochemistry, it is logical to focus on planets which have the propensity to harbor surface water in liquid-phase.  Thus we would want to know how frequent rocky Earth-like  planets occur within a radial region known as the habitable zone\footnote{The habitable zone is defined as the orbital radius around a star of a given spectral type that would provide a surface temperature such that water, if present, would be in liquid form (as opposed to solid ice or water vapor).  The classic reference for this term is by \citet{Kasting93}.}.    Over the past decade, with significant contributions from the first successful planet hunters \citep{mayor95, marcy96}, the search for extra-solar planets has become a dominant field in astronomy.   Quite recently NASA's Kepler satellite provided the first clear evidence of rocky planets and we now have a few candidate objects that reside within the habitable zone \citep[e.g.,][]{borucki12}.

Of course the mere detection of a rocky world, which is inferred to have an environment suitable for surface water, does not conclusively state that reproductive life is present in any form.   
To answer this question, there are two main perspectives one can take.  The first of which is to study the detailed history of life on Earth over the last 4.6 billion years of planetary evolution, i.e., an "inside-out" perspective.  The second "outside-in" perspective is guided by extra-solar clues from planetary systems in formation, where we seek to know how the earliest evolutionary stages fostered an environment where liquid water and organic monomers\footnote{Definition: ``a molecule of any of a class of compounds, mostly organic, that can react with other molecules of the same or other compound to form very large molecules, or polymers. The essential feature of a monomer is polyfunctionality, the capacity to form chemical bonds to at least two other monomer molecules. Bifunctional monomers can form only linear, chainlike polymers, but monomers of higher functionality yield cross-linked, network polymeric products.'' From Encyclopedia Britannica Online, s. v. "monomer," accessed July 26, 2013, http://www.britannica.com/EBchecked/topic/389906/monomer.} were present and were allowed to react towards chemical complexity. 
In a sense this focus is on the initial chemical and physical conditions associated with young so-called ``proto-planets''.
 
 The latter approach is not without caveats.  One such caveat is that what we categorize astronomically does not necessarily represent the physical or chemical conditions on the surface of forming worlds.
Rather, we study the conditions of gas and solid particles that precede planetary birth.   Thus observations of the pre-planetary materials can be compared to the record of materials in our solar system.   In this paper, we approach the topic from both perspectives.
 First, we summarize the clues from our own history by looking backwards.  Thus we explore this history of life on the Earth and the process of planet formation by the study of our own system with a basic physical/chemical framework.      Second, we outline the astronomical perspective where we can look forwards by observing systems in various stages of formation.  This discussion is by no means a review of the field of origins research and the reference list will be vastly incomplete.   Rather the references chosen are intended to be illustrative and to provide a reference trail for a young researcher.  

\section{Looking Backwards}

\subsection{History of Our Planet and Life}

\subsubsection{Atoms and Molecules}
\begin{figure}
\caption{Percentage of mass of the human body dominated by various atoms.  The atoms comprised in ``other'' are Phosphorus, Potassium, Sulfur, Sodium, Chlorine, and Magnesium and other trace elements.}
\includegraphics[height=.4\textheight]{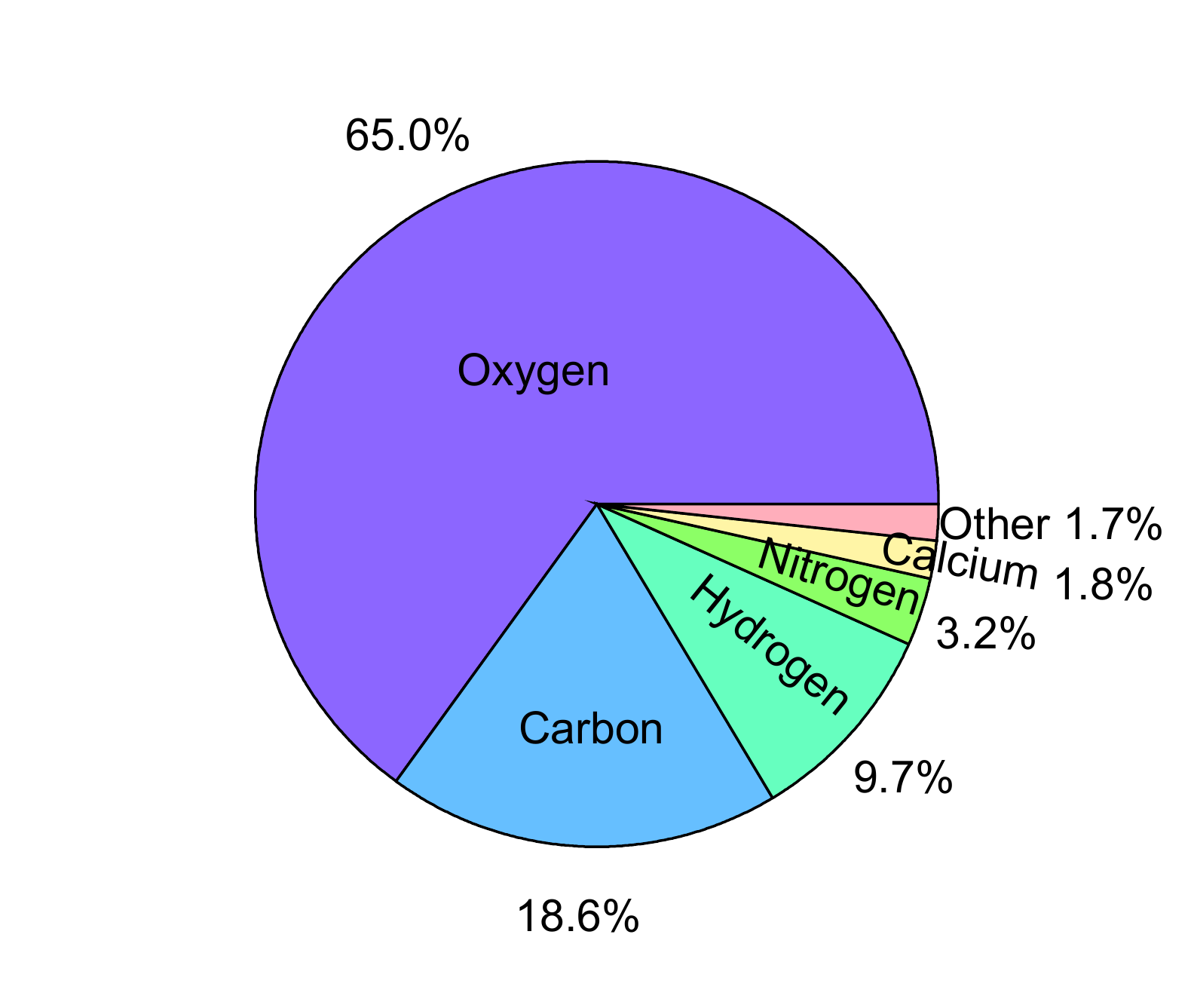}
\label{fig:atoms}
\end{figure}

Before exploring the history of our planet and clues to the timing of life's origin, it is useful to provide context.   In Fig.~\ref{fig:atoms} we illustrate the percentage of human body mass that is comprised of various atoms.  Living mass is dominated to the level of 96\% by the most abundant non-inert (i.e. reactive) elements: H, O, C, and N (hereafter CHON).   The dominance of oxygen is due to the fact that we are mostly comprised of water, while carbon, with its potential to share 4 bonds, provides the backbone for the chemistry of life.     One might speculate whether this will always be the case for life, with water providing the liquid medium and carbon the basis for biochemistry.   For example, silicon has the same number of bonds as carbon.  However,  C readily bonds with itself, while Si bonds primarily with oxygen.
Thus, in space there is a demonstrated dominance of the gas and ice chemistry to make carbon-based molecules and water \citep{cc12}.   These species are labelled as volatiles as they readily evaporate into the gas from the solid state.   This is opposed to refractory components, such as silicon-oxides, that generally exist as solid minerals.
Thus the initial state of the system will not have much free silicon available.   For an interesting discussion in this regard the reader is referred to \citet{benner10}.   

\begin{figure}
\caption{The chemical composition and structure of DNA (Credit: Madeleine Price Ball).}
\includegraphics[height=.4\textheight]{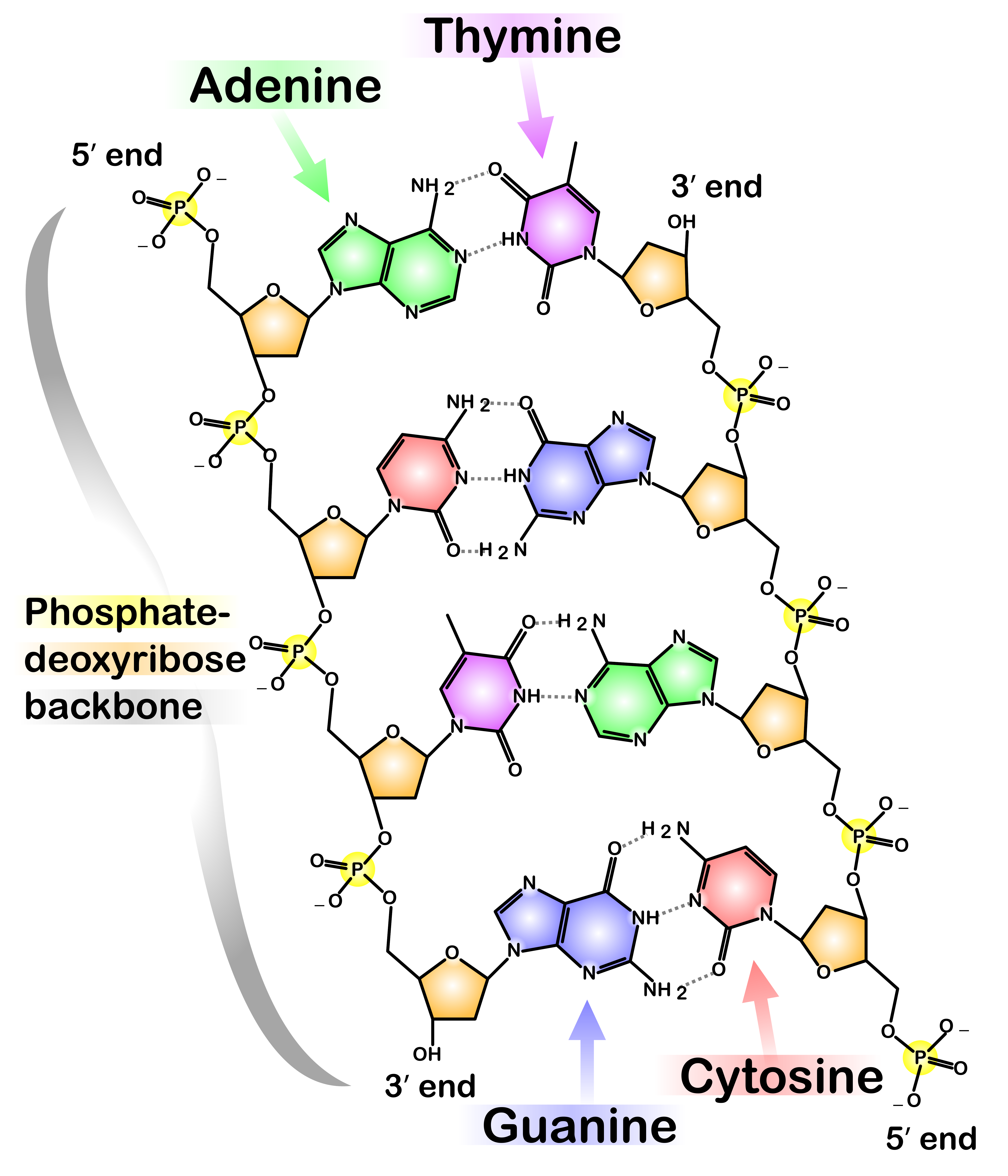}
\label{fig:dna}
\end{figure}

Another perspective is that all life on Earth uses water as the liquid medium to facilitate and participate in biological chemistry.  All life also uses deoxyribonucleic acid  (DNA) to store and pass on genetic information.  This points towards a single common origin for life on Earth.   Fig.~\ref{fig:dna}  illustrates the structure and composition of the DNA molecule.   Labeled in this figure are the four DNA bases whose pairing store genetic information (adenine, guanine, cytosine, and thymine) and the phosphate-sugar (deoxyribose) backbone.  From an astronomical perspective the complexity of this molecule is astounding.  
As we will discuss in the following section, the chemistry that precedes and is associated with planetary formation readily produces complex molecular species.  However, the most complex molecules involving more than one element detected and identified in space have only $\sim 10-20$ atoms, compared to the hundreds in DNA.   A host of amino acids, the pre-cursors to proteins, are found in some meteorites \citep[e.g.,][]{pizzarello10}.   But, even in that case, for the pre-life materials
only monomers are sampled.  These monomers must be concentrated in some fashion and provided a source of chemical/geothermal energy to form polymers.  This would be the initial, and highly uncertain, steps in the direction of self-replicating chemistry.  This clearly happened on the planet itself, perhaps aided or jump started by the supply of pre-existing ``pre-biotic'' material during birth, which we address below.   
For greater discussion see \citet{Orgel98} and the numerous articles compiled by \citet{origins_csh}.

\subsubsection{Radiometric Dating and the Age of the Earth}

One important question regarding life's origins is timing; how quickly did life originate in demonstrative ways in the fossil record?  Here the term fossil is used loosely, as the record of ancient life is not comparable, or as easy to isolate, as for vertebrate life-forms in the post-Cambrian epoch (< 0.58 Billion years - hereafter, Byr).    The general technique is to separate the fossil evidence from  absolute dating and instead 
use radiometric dating to determine the age of surrounding rocks.   Radiometric dating  relies on the measurable decay of an unstable isotope into stable form over a defined timescale.   The decay rate is well calibrated in the laboratory with measurements of the half-life, which is the time over which 1/2 of the original parent is transferred into the daughter product. 

One useful method to date ancient materials, which can be used as an example, is the U-Th-Pb chain, which has three independent parent/daughter products, each with a long half-life.   The usefulness of this is that each of these can be measured independently and, with agreement, the age has greater reliability.    This is only one example and there are numerous other parent/daughter pairs and combinations.   More explicit discussions of the methodology used in geochemistry can be found in the books by  \citep{Faure05} and \citet{Allegre08}.
The U-Th-Pb chains are given below:

\begin{enumerate}
\item $^{238}$U $\rightarrow ^{206}$Pb (half-life of 4.4683 Billion years)
\item $^{235}$U $\rightarrow ^{207}$Pb (half-life of 0.70381 Billion years)
\item $^{232}$Th $\rightarrow ^{208}$Pb (half-life of 14.0101 Billion years).
\end{enumerate}  

\noindent Half-lives for this system taken from \citet{Allegre08}.  In a rock the measurement of each of the isotopes can be used in the following set of equations \citep{Faure05}:

\begin{equation}
\frac{^{206}Pb}{^{204}Pb} = \left(\frac{^{206}Pb}{^{204}Pb}\right)_i + \frac{^{238}U}{^{204}Pb}(e^{\lambda_1 t} - 1), \\
\end{equation}

\begin{equation}
\frac{^{207}Pb}{^{204}Pb} = \left(\frac{^{207}Pb}{^{204}Pb}\right)_i + \frac{^{235}U}{^{204}Pb}(e^{\lambda_2 t} - 1), \\
\end{equation}

\begin{equation}
\frac{^{208}Pb}{^{204}Pb} = \left(\frac{^{208}Pb}{^{204}Pb}\right)_i + \frac{^{232}Th}{^{204}Pb}(e^{\lambda_3 t} - 1). \\
\end{equation}

\noindent These equations are normalized to the stable isotope $^{204}$Pb, which is not radiogenic.  The subscript $i$ refers to the initial isotopic ratio in the rock when it formed.   $\lambda_{1,2,3}$ refers to the half-lives given above.
If you measure the isotopic compositions of the rock using techniques such as mass spectrometry \citep{Nier41} and plot (for example) $\frac{^{206}Pb}{^{204}Pb}$ against $\frac{^{238}U}{^{204}Pb}$ then the slope of the line will contain both the half-life and the time of decay.  Thus time can be measured.
   Key points are as follows.  Radioactive dating can be measured reliably, provided the system is closed throughout its history.  That is, there is no isotopic exchange as a result of, for example, diffusion or metamorphism.   The measured decay time refers to the time since the system closed, i.e. the rock formed.   One key question regards the initial composition which is uncertain.    One method to get around this uncertainty is to use minerals that exclude certain atomic species upon formation.  For example, zircon (ZrSiO$_4$) contains very little Pb at the time of formation.  Thus the daughter product will be unique.    

From the perspective of dating ancient materials, an important problem is that the Earth is geologically active and it can be difficult to always isolate specific minerals that exclude Pb during formation (again as one example).   
The first reliable estimate of the Earth's age came from Clair Patterson \citep{Patterson55} who used meteorites as the proxy. His measurement determined the age of a sample of meteorites,  to be $\sim 4.55$~Byr and also provided primordial isotopic ratios.    Assuming that the Earth and meteorites formed from the same isotopically well mixed material, this then unlocked the capability to explore the question of the Earth system.   Clearly Patterson's age is a reference point for the age of the first solids in the solar system, which we now know to very high accuracy,  4564.7 $\pm$ 0.6 million years ago \citep{amelin02}.  The formation of the Earth itself likely occurred slightly later ($\sim$tens of millions of years) and the reader is referred to the following reference by \citet{Allegre95}.

\subsubsection{Early Evidence for Life}
 
In terms of the question of when life began the following references \citep{Orgel98, gk99, ns01, lopez-garcia06, javaux10rev, arndt12, Bosak13} can be used to explore the question more deeply.  In this review a few salient points are summarized.     Below we assume that the Earth was born 4.6 Byr ago.   For a framework the geological history of the Earth is divided into four eons:  (1) Hadean - 4.6 to 3.8 Byr: this is the hellish Earth as the young planet was forming, cooling, and repeatedly bombarded by large and small planetesimals during the time of planet formation;  (2) Archean - 3.8  to 2.5 Byr: the phase of early life, when the first plant fossils are found and encompasses the stage when life was beginning to take root on the planet; (3) Proterozoic - 2.5 to 0.57 Byr: stage where life began to dominate our planet, changing for example the atmospheric composition \citep{Bekker04};
(4) Phanerozoic - 0.57 Byr to present: the stage of visible life.

 There are 3 lines of evidence that life arose rather quickly on the surface of our planet during the Archean eon about 3.5 Byr  and perhaps as early as 3.8 Byr.  These are briefly outlined below.
 
\begin{enumerate}
\item $^{12}$C  and $^{13}$C ratios:  living tissue metabolizes $^{12}$C more easily than $^{13}$C.  Thus the chemistry of life leads to an excess of $^{12}$C relative to $^{13}$C when compared to inorganic carbon.  In some ancient rocks in Greenland (3.8 Byr) there is evidence for this excess in some carbonaceous inclusions inside phosphate minerals \citep{Mojzsis96}; in one instance the isotopically light carbon is in small graphitic globules potentially attributed towards ancient Archean plankton \citep{Rosing99}.  However, there is some controversy as the rocks towards one site have been claimed to be igneous (melted), which would not preserve the signature \citep{Fedo02, Moorbath05}; but see also \citet{grassineau06}.

\item Stromatolites: stromatolites are colonies of layered bacteria that are found today in Shark Bay, Australia.  Sedimentary rocks are common on the Earth and in the fossil record stromatolites in a sense would be organosedimentary layered structures \citep{buick81}.     The strongest case for ancient stromatolites appears at 3.43 Byr \citep{walter80, lowe80, Hofmann99} and 3.416 Byr \citep{Tice04}.  At these sites multiple types (morphological) of  stromatolites are found \citep{Allwood06} in  potentially shallow water environment to foster formation \citep{Lowe83, Allwood07}.   As always there are some uncertainties and there are other ancient sites; summaries of the state-of-the-art can be found in \citet{Tice11} and \citet{Bosak13}.

\item Microfossils: \citet{Schopf87} and \citet{schopf93} presented the first evidence for microscopic filamentary carbonaceous structures detected in 3.5 Byr old rocks, which are posited as ancient cyanobacteria that formed in  shallow water.    Over the past decade this claim has been called into question with the statement that the rocks are likely samples of a hydrothermal vent system \citep{Brasier02}.   The microscopic structures in question were argued to be graphitic artifacts deposited in this environment.   More recently \citet{Marshall11} found that similar structures from the same site were instead a series of fractures filled with the mineral haematite; carbonaceous material is found the immediate vicinity.  They argue that biology is not ruled out as the origin of the carbon, but the microscopic structures themselves are not evidence for biology.   \citet{Schopf11} posit a counter-argument that the original fossils were not examined.    
However, there remains compelling microfossil evidence for life in the early Archean (3.2 and 3.4 Byr old rocks).  Here the microscopic structures seen exhibit inner hollow cavities with carbonaceous (cell?) walls \citep{Javaux10, Wacey11}.  At the older site,  the carbonaceous material is isotopically light, which links geochemical with the morphological evidence \citep{Wacey11}.  
\end{enumerate}

\subsubsection{The Young Earth}

In the preceding paragraphs, we summarize the evidence that life existed on Earth at least $\sim$3.5 Byr ago and potentially earlier.   Furthermore, another important point is that  geochemical  evidence from ancient zircon grains  suggests that Earth had oceans of water up to 4.4 Byr ago \citep{mojzsis01, wilde01}.  Thus it appears that water was on the surface of our planet during very early stages.

Based on the crater-coated lunar surface, we also know that the early history of the solar system was a violent time.    Since the Earth has an effective  cross-section that is 20 times larger than the Moon it was hit 20 times more.  In fact, the formation of the Earth was not a quick process, but over the first few hundred million years we suffered from constant bombardment from space.   Some of these impacts would have been enough to vaporize any young oceans on the surface, perhaps sterilizing incipient life \citep{ns01, sleep2010}.  Determining the history of these impacts requires high fidelity images of the Moon and counting the number of craters with different sizes.   A portion of the Moon that has a large number of craters, with impacts piled on impacts, is older than a surface that has less crater coverage.   This provides only relative ages as opposed to absolute values.
  Fortunately one of the many benefits of the Apollo missions was a sample of Lunar rocks from sites with different crater densities.  These rocks can be aged via radiometric techniques.

\begin{figure}
\caption{The underlying plot shows the number of craters greater than 1 km in size per km$^{2}$ in a site with a given age noted by the lunar mission and/or crater name.   Plot is a recreation of that shown in \citet{Neukum01} using their Eqn. (5) to delineate the fit to the data of crater densities and radiometric ages from \citet{sr01}.  The geological eons are shown on the bottom, along with potential timing of early cellular life and presence of surface water.}
\includegraphics[height=.6\textheight]{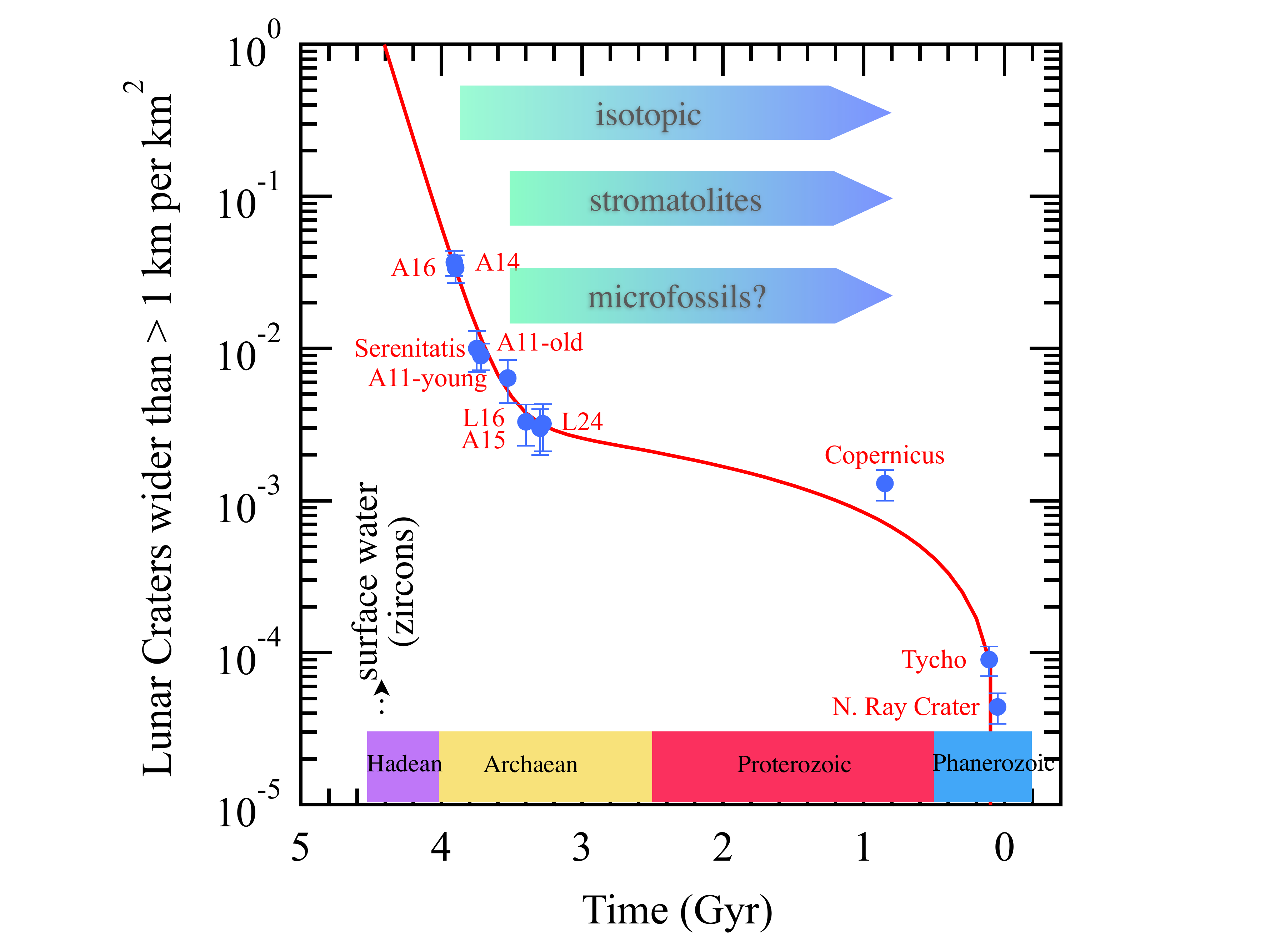}
\label{fig:lunar-life}
\end{figure}

A sample of these results is given in Fig.~\ref{fig:lunar-life}, where we reproduce the chronology of impacts in the solar system along with the history of life described earlier.  A few things are notable: in the plot there is a strong decay in lunar impacts with time, 3 orders of magnitude in crater density by $\sim 3.8$~Byr.   This means only 5\% of the craters are younger than 3 Byr \citep{Neukum01}.  There is a leveling off in the plot at a y-axis value of 10$^{-3}$ (units of craters wider than 1 km per km$^{2}$) which is not necessarily indicative of a time of constant crater rate, but rather more of a statement about the stochastic nature of the process.    In this light, the general decrease in time during the Hadean/early-Archean may not be a smooth decline, but one punctuated by periods of intense impact and others with much less with an overall decay in time \citep[see discussion in][]{Zahnle06}.    Based on Fig.~\ref{fig:lunar-life}, the geological evidence points toward the emergence of life on our planet as the frequency of energetic impacts decayed.   \citet{Southam07} argues that the Earth was essentially habitable throughout the Hadean, while \citet{sleep2010} makes an argument for photosynthesis as early as 3.8 Byr.   The important point here is not the details, which are complex, but that simple life did not take billions of years to take root.  Life potentially began even as the Earth was still subject to impact events of significantly higher energy \citep{Zahnle06} than the Cretaceous-Tertiary extinction event \citep{Alvarez80} that led to the extinction of the dinosaurs.

\subsubsection{CHON in the Context of Planetary Birth}

The formation of the Earth involved the gathering of rocky materials, during a period of accumulation and accretion.  In this context an important question is how did the Earth receive carbon, oxygen, hydrogen, and nitrogen.    We know that the rocks of the Earth are mostly comprised silicate minerals.  In fact, the total amount of carbon in the Earth's mantle is $\sim 10^{23}$ g \citep{dh10}, which is a tiny fraction of the Earth's mass ($\sim 0.002$\%)\footnote{We note that significant amounts of carbon could still be present in the Earth's core \citep{dh10}.}.  Similarly, as a relatively water-rich terrestrial planet, the Earth's water content by mass is sparse.  \citet{Marty12} estimates that the Earth as a whole contains 7 ``oceans'' of water (where ocean is defined here as the surface water), which again is only a small fraction, 0.2\% of our planet, by mass.   Clearly these materials were provided, but not in bulk form along with silicates.   

To explore this question we must understand the context of planet formation.  Silicate solids are formed in interstellar space  with microscopic, average $\sim 0.1 \mu$m, sizes \citep{draine03}.     
These small dust grains are then delivered to planet-forming disks.
   These tiny grains, called dust in astronomy, are the seeds of terrestrial worlds.  The details of the various stages of planet formation can be found in the reviews by \citet{yk13} and \citet{Morby12}, along with the book by \citet{Armitage10}.  Within a few million years the tiny grains must coagulate and grow in the disk into planetesimals ($\sim$1 km), until gravity can take over.   This is not a simple growth process and there are a number of uncertain steps along the way.     During the next stage these the planetesimals grow to planetary embryos (Lunar to Mars sized) under mutual gravitational interactions where the larger bodies grow disproportionately.     The next stage involves these embryos traversing a sea of planetesimals.  This is the phase of truly giant impacts (formation of the Moon \citep{Benz87}) and gradual accumulation to Earth/Venus sizes.

\begin{figure}
\caption{Approximate sequence of equilibrium condensation drawn from the information given by \citet[][see Table~IV.7]{lewis04}.   The white boxes show the approximate transitions of various elements into minerals and ices 
from the gaseous state assuming solar nebula disk pressures (10$^{-2}$ bar).  While the blue boxes provide the formation sequence as gas cools (see text for discussion of condensation vs. evaporation).   Some modifications include a question as to whether N is found as NH$_3$ or as N$_2$ or even N, as the latter two have lower condensation temperatures.  Thus a range has been shown.   The complete sequence given in \citet{lewis04} is shortened. The green dashed line roughly delineates the line between refractory solids and ices, while the yellow line notes species that remain in the gas.  Note that C and N straddle this line.  O in the form of H$_2$O will also be found both in the gas and ice at various radial positions in a protoplanetary disk.  A fraction of C, H, O, and N will also be bound in organics which, depending on the form, have condensation/sublimation temperatures in the range of 50 -- 150~K \citep{Oberg09a}. }
\includegraphics[height=.45\textheight]{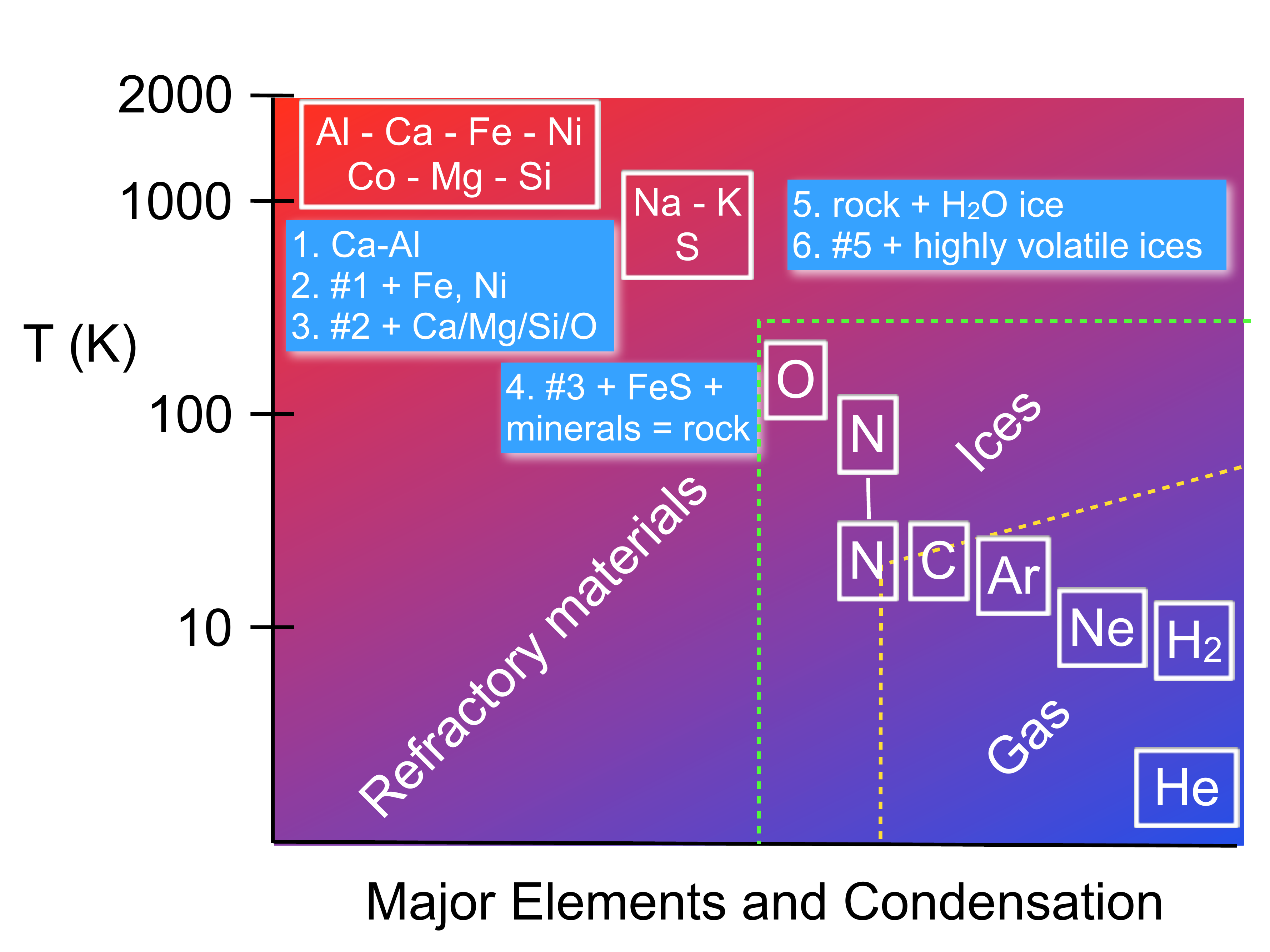}
\label{fig:cseq}
\end{figure}

 One school of thought, from the cosmochemical/meteoritic community, posits that the solar system at the earliest stages was very hot ($> 1400$~K) such that all atoms are in the gas \citep[e.g.][]{grossman72, wh93, eb00}.  As the nebula cooled, solids formed a sequence of condensation.  The formation of planets then proceeds through the stages outlined above.  As we will discuss below this is not the entire picture, but it nicely provides an illustrative example of the chemical physics that is involved in incorporating CHON into terrestrial worlds.   Fig.~\ref{fig:cseq} provides a graphical illustration of how this  proceeds loosely based on the book by \citet{lewis04}.  As the gas cools in thermodynamic equilibrium the most refractory materials with the highest condensation temperature\footnote{We note that the pressure in a young disk is too low for the liquid phase to exist.  Thus the phase transition is between the vapor and solid states. However, liquid water was most certainly present inside the largest planetesimals. } forms the first minerals (\#1).  In this case Ca and Al materials would be the first solids, a fact that is consistent with Calcium Aluminum-rich inclusions being the oldest material found in the solar system \citep{amelin02}.  At lower temperatures ($\sim 800-1200$~K) magnesium/iron-rich silicates would form what are essentially the solar system rocks (\#2-4).  Lower temperatures ($<$ 150--200~K) are required for water ice to condense (\#5), and even lower temperatures ($\sim$20~K; \#6) for molecular nitrogen and carbon monoxide (two key carriers of C and N).     These elements are also incorporated into organics which condense at higher temperatures, comparable to water.
 Finally we have the materials that always remain in the gas: noble gases, molecular hydrogen, and helium.   
 
The process of evaporation (loss of material as opposed to formation) would work in reverse, thus this is not necessarily a temporal sequence \citep{davis_messii}.   Moreover the isolation of grains in meteorites whose origin precedes the solar system \citep{zinner98} requires the presence of unaltered or unevaporated material.  Some molecular signatures (deuterium enrichments in water ice and organics) also record evidence of chemical processes not in thermodynamic equilbrium \citep{robert_messii, willacy09}.     Regardless Fig.~\ref{fig:cseq} is useful in terms of outlining that different solids condense from the gas to the solid at different temperatures.
 CHON molecules that are important to life are among the most ``volatile'' and require temperatures $\sim$100~K or lower for incorporation.

Let's now examine what happens from the other perspective where grains are formed in the interstellar medium and supplied to the disk via collapse along with volatiles perhaps in the gas or even as ice coatings on the dust.   Going back to our physical understanding of planet formation, it is during the early phases
when the grains are microscopic, they have the largest surface area and thus greater exposure to collisions with a CHON molecule dominated gas.   Moreover we need temperatures below 100 K. 
Based on our understanding, to be discussed later, this can occur prior to stellar birth or in the outermost reaches of the planet-forming disk.     In these phases CHON atoms coat the silicate solids as molecular ices (e.g. pre-cometary ices).  This includes both water and organics that are readily detected as gases in cometary coma \citep{mc11}.   The comet forming zone resides somewhere near the location of the giant planets at $>$5--10 AU from a star like the Sun.

To investigate what happens at 1  AU, if we take a bare 0.1 $\mu$m dust grain we can can balance the energy absorbed with the energy emitted to determine expected temperature.    The energy absorbed ($E_{abs}$) and emitted ($E_{em}$) are:

\begin{equation}
E_{abs} = F_{\odot} ({\rm 1\;AU}) Q_{abs} \pi a_{gr}^2,
\end{equation}

\begin{equation}
E_{em} = Q_{em} 4\pi a_{gr}^2 \sigma T_{gr}^4.
\end{equation}

\noindent Here $F_{\odot}$ is the solar flux at 1 AU,  $a_{gr}$ is the grain radius, $T_{gr}$ the grain temperature, and $\sigma$ is the Stefan-Boltzamann constant.  $Q_{abs}$ and $Q_{em}$ are factors that describe the physics of the interaction of the small particles with radiation, with the absorption occurring where most of the solar energy is emitted (visible light) and emission at longer infrared wavelengths (heat).   It turns out that 0.1 $\mu$m astrophysical solids are good absorbers of visible light, but inefficient emitters; typical numbers that can be assumed are $Q_{abs} = 1$ while $Q_{em} \sim 2 \times 10^{-3} (a_{gr}/1\;\mu {\rm m})T_{gr}$ \citep{Draine81}.  Putting in these parameters we see that the typical grain temperature at 1 AU will be hundreds of K.  

Thus the initial rocks out which the Earth was made would not have water ice.  This result is simplistic as it assumes a bare grain sitting alone at 1 AU with approximate grain properties.  However, more detailed radiation transfer calculations confirm the general picture \citep{Podolak04, Lecar06, Davis07, Dodson-Robinson09}.  This well known result suggests that the solar nebular disk, and extra-solar protoplanetary disks, should have what is called a ``snow-line''.  Inside this line water (and most organics) exist as gaseous vapor and as ices further from the star.    The composition of bodies in the solar system clearly reflects this.  There is an inferred gradient of water content in the asteroid belt from meteoritic samples \citep{Morby12} and comets, the most ice-rich bodies by mass, formed far from the Sun. 

 \subsubsection{CHON in Meteorites and Comets}
 Asteroids and comets represent remnant, perhaps primordial, material left over from the birth of our solar system.  As such they offer the greatest opportunity to explore the chemical/physical state of our solar system at birth.  This record has been lost on our geologically active planet.

  Asteroids are sampled most directly by the study of meteorites. 
 There are a number of meteoritic classes, based, in part, on whether the material is undifferentiated (and therefore more primitive), or differentiated (i.e. processed) \citep{krotmcp}.  
   Unequilibrated meteorites thus provide the clearest record of the starting materials and are composed of  (1) chondrules, small ($\sim$mm-sized) spherical igneous silicate particles that are believed to have been transiently heated; (2) CAI's, non-igneous calcium aluminum rich inclusions, the oldest minerals in the solar system, (3) a matrix of fine grain, micron-sized, particulate matter that exists as the glue between these components.     The matrix is comprised mostly of silicates, but often contains hydrated silicates and carbon.  Thus many planetesimals contained water and have evidence of aqueous alteration.    
 
  The most primitive meteorites are the CI chondrites, which have the lowest amount of chondrules, comprised mostly of matrix \citep{Krot09}.
     The bulk abundance of CI chondrites mirrors that of the Sun for all but the most volatile species \citep{lodders03}.     Organics are found in both insoluble form (insoluble organic matter) and water-soluble (soluble organic matter).
 In one CI  meteorite alone (Murchison) over 70\% of the carbon is found in the insoluble portion in macromolecular form ($\sim$ 100 carbon atoms  plus H, O, N).  In the soluble portion $\sim 30$\% of the carbon is in over 1000 molecular species including 100 amino acids (terrestrial proteins are made of only 20 amino acids) \citep{pizzarello_messii, pizzarello10_csh}.  Furthermore, nucleobases are also detected in several different samples \citep{Martins08, Callahan11}.     The amount of soluble and insoluble organics ranges within and between meteorite classes \citep{pizzarello_messii}.

 \begin{figure}
\caption{Figure showing the D/H ratios measured for a variety of solar system and interstellar objects in hydrogen, water, and
some organics.   Values taken from the following references, hydrogen \citep{Lellouch01, mahaffy98, gg98, Feuchtgruber13}, water \citep{Greenwood11, Alexander12, mc11, Hartogh11, Persson13, Taquet13, Parise03}, organics \citep[][]{Wang05, ph05, pizzarello_messii, meier_dcn, Roueff03, Parise06, neill13b}.
IOM refers to the insoluble organic material in meteorites, while SOM refers to the soluble organic material; ISM refers to the interstellar medium in this case dense star-forming gas.   }
\includegraphics[height=.45\textheight]{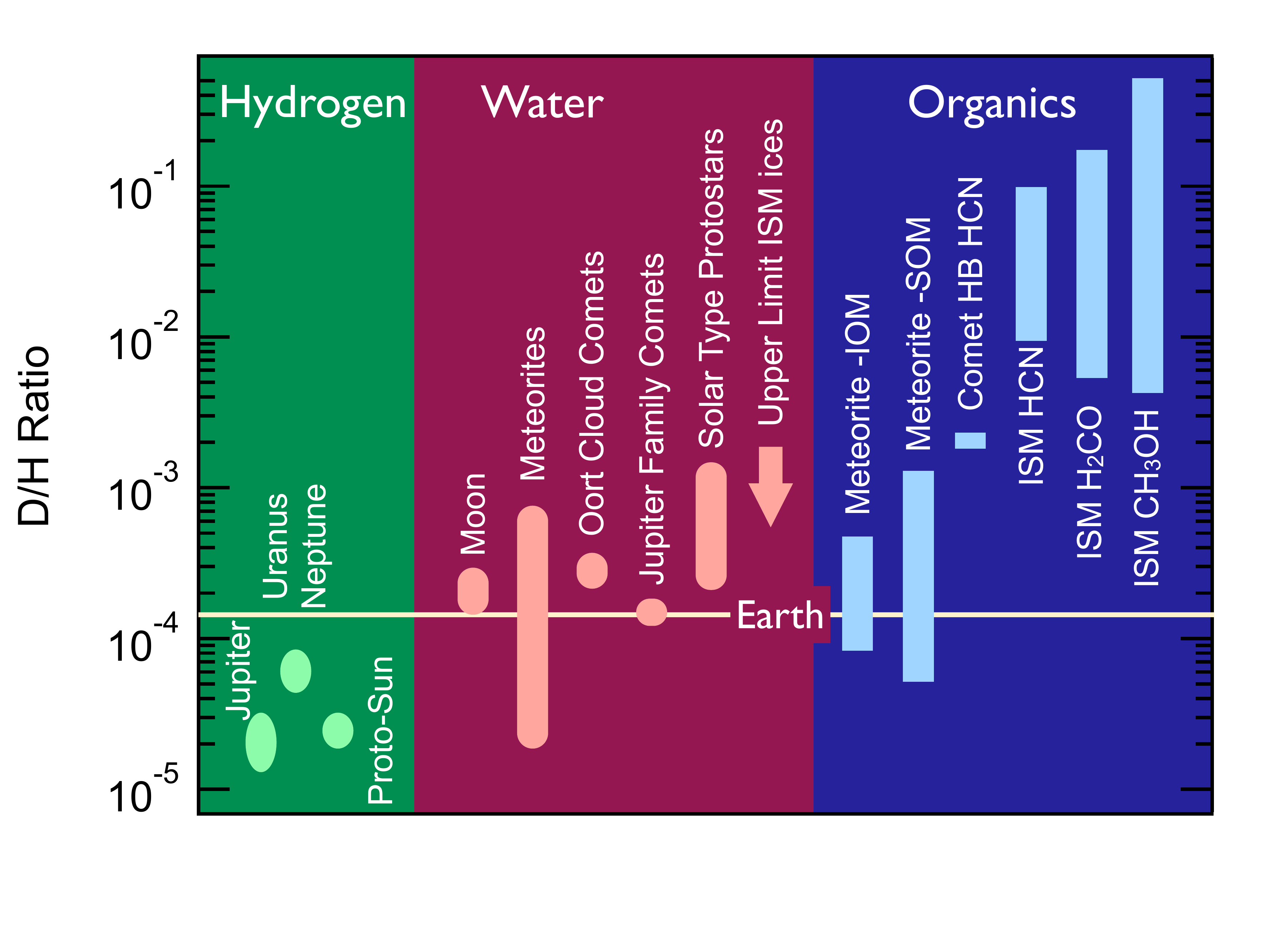}
\label{fig:dh}
\end{figure}

Comets are comprised of both ices and solids and the composition of cometary ices contains some of the most volatile molecular material (e.g. CO$_2$, CO) available in the solar system.     Thus comets also represent potentially pristine remnants of solar system formation.
Long period comets, with periods $>$ 200 yrs, originate in the Oort cloud, while short period comets
($<$ 200 yrs) formed in the Kuiper belt \citep{Levison97}.   Oort cloud comets are argued to originate in proximity to gas/ice giant planets and scattered to distant orbits via planet-comet interactions \citep{f97, dwld04}.    
Water is the most abundant cometary volatile with trace amounts of CO$_2$, CO, and other species.
When looking broadly at both abundant and trace compounds there is a gross similarity of the volatiles seen in the evaporated cometary coma and that of interstellar ices  \citep[e.g.][]{bm00, ecw04}.    However, CAI's and crystalline silicates \citep{2006Sci...314.1735Z} have been detected in the sample return of the Stardust mission.  These both require very high temperatures $>$ 1000~K to be mixed with ices that condense near 20~K.  
The presence of essentially asteroidal material in comets is strong evidence for large-scale radial mixing in the solar nebula disk \citep{brownlee_stardust}.
 Moreover, there are differences in the relative molecular compositions as some comets that originate from the Oort cloud and Kuiper Belt are depleted in organics, hinting at gradients or mixing into and within the comet forming zones \citep{mc11}.

Both the soluble and insoluble portion of the meteoritic matrix exhibit enrichments in the heavier stable isotopes of C, H, and N \citep{pizzarello_messii}.   Comets also contain isotopic enhancements in H and N \citep{mc11}.
  In this review we focus on deuterium, solely as an example, but the isotopic enrichments seen in C and N provide useful constraints as well. 
   Deuterium was created in the big-bang and the D/H of hydrogen is measured to be $1.5 \pm 0.1 \times 10^{-5}$ in the local solar neighborhood \citep{linsky98}.    The process of isotopic fractionation comes about because of the difference in the zero
     point energy among hydride isotopes  is quite small ($\sim 200 - 500$~K), but nonzero.    At low temperatures ($< 100$~K) non-equilibrium gas phase  reactions involving, for example HD, leads to a preferential transfer of the D as opposed to the H; leading to (D/H)$_{\rm X}$ $>$ (D/H)$_{\rm H_2}$.      In Fig.~\ref{fig:dh} we show the D/H ratios measured in hydrogen, water, and organics in a variety of objects in the solar system and interstellar medium.
     We will discuss the full import of this plot when we discuss the chemistry of the interstellar medium; however, one key statement is that Earth's ocean water has about an order of magnitude enrichment relative to the main reservoir of deuterium, hydrogen in the Sun (i.e. proto-Sun in Fig.~\ref{fig:dh}).        Comets are well known to have deuterium  enrichments in both water and, in one measurement, HCN.  
  Organics and water in meteorites also have enrichments in both the soluble and insoluble organic molecules.   In some instances, the D/H ratio of water in solar system bodies is similar to that measured in Earth's oceans.  Thus the D/H ratio might be a fingerprint tracing the origin of water on our planet.   This is for some meteoritic classes \citep{Alexander12} and perhaps Jupiter family comets, although the sample is very small \citep[][]{Hartogh11}.  At face value this might hint that sources for Earth's water are found beyond our orbit in the solar system.

 \subsubsection{The Delivery CHON to the Earth}

The current consensus is the that pre-Earth materials at 1 AU would not have water or organics present, but see \citep{murali08, king10} for an alternate view.     Thus either the Earth (1) obtained key volatiles from the gas or (2) received some supply from beyond the so-called snow-line as part of its formation.    

\begin{enumerate}
\item It is a general consensus that the Earth has had two atmospheres.    During the first $\sim$few Myr of evolution as the Earth formed it was surrounded by H$_2$-rich nebula gas.   As the proto-Earth grew in size it could gravitationally capture gas from the nebula \citep{Hayashi79}, producing the first - H$_2$ rich - atmosphere.  
Nebular gas capture is a strong candidate for how the Earth obtained its inventory of the most volatile gases, He and Ne \citep[e.g.][along with any addition from the solar wind]{Harper96, Porcelli03}.     The young Earth would have been continually bombarded by impacts during this time frame and these impacts would have lead much or part of the surface to be covered by a silicate magma ocean \citep{Elkins-Tanton12}.  Reactions between this molten rock and the H$_2$ gas could produce water vapor \citep{Ikoma06}.    Thus in this theory the Earth would have gained its water from solar nebular disk H$_2$ dominated gas.   

However there are some issues with respect to the timescales involved in terrestrial planet formation and the lifetime of the gas-rich nebula.   Based on Hf-W chronometry it is estimated that the Earth's core formed on timescale of $\sim 30$~Myr \citep{yin02, kleine02, schoenberg02, touboul12}, while current estimates of the half-life of gaseous disks is $\sim 2$~Myr \citep{williams_araa}.     Thus the gas content of the Solar nebular disk at the time Earth reached its largest gravitational potential is uncertain.     Stronger evidence comes from the isotopic ratios of noble gases, which suggest that the primordial or first atmosphere was lost via hydrodynamic escape \citep{Porcelli03, Zahnle07, Holland09}.  Thus, it is thought that the Earth lost much of the initial atmosphere and generated its second atmosphere by outgassing, which then reflects a composition resulting from impact delivery along with any atmospheric loss \citep{Morby12, Halliday2013}.   

\item Numerous models of  terrestrial planet formation suggest that during the early stages the material accreted in terrestrial planets originates in relative close proximity.  However, the loss of nebular gas during the stage where embryos and planetesimals undergo mutual interactions leads to strong perturbations and collisional growth from material over a larger range of distances \citep{Morby12}.    In this case the delivery of water and volatiles can readily occur via the transport of planetesimals that formed beyond the snow-line to the young proto-Earth \citep{rql04, obml06}.   Key facets of these models are the timing of the formation of giant planets, their orbital eccentricities, orbital migration, and the overall evolving planetary system architecture.     At present models suggest that the most likely source for planetesimal delivery are radii that correspond to today's asteroid belt \citep{rql04, obml06}, i.e. water-laden rocks as opposed to comets.  One intriguing area of new research is the possible detection of comets in the outer parts of the main asteroid belt \citep{Hsieh06}, which represent a yet to be characterized potential source term.   
\end{enumerate}

 \subsection{An ``Astro'' in Astrobiology}
In the previous section we have explored the question of life's origin in a broad sense.   We have outlined the various lines of evidence that life started rather quickly on the planet's surface, perhaps while the Earth was still undergoing bombardment.    These impacts appear to be crucial in terms of the delivery of CHON volatile material to our young world.  Of course this assumes that this volatile content is provided to the Earth and not destroyed \citep{Ahrens89, cs92}.  Thus one part of the ``astro-'' in astrobiology is the question of how these volatiles came to be implanted in the planetesimal impactors in the first place.     We are in a sense thus divorcing ourselves from a question of specificity in terms of the
mechanisms by which key prebiotic organics produced life, which is rightly a question for biochemists and chemists \citep[e.g.][]{benner10_cshp, sutherland10_cshp}.    As we will outline below, there is a wide variety of crossover between species detected in meteorites, comets and interstellar space.  Indeed, some of these molecules of interstellar origin might have been useful in jump starting the chemistry of life.
Therefore we now address the more general question of how to create the molecules that we measure in comets and asteroids, the final samples of the primordial solar system.

\section{Looking Forwards in Time:  The Birth of Solar Systems}

\subsection{The Interstellar Medium and Stages of Star Formation}

\begin{figure}
\caption{Schematic of the various phases of star and disk formation, and their estimated time-scales.}
\includegraphics[height=0.5\textheight]{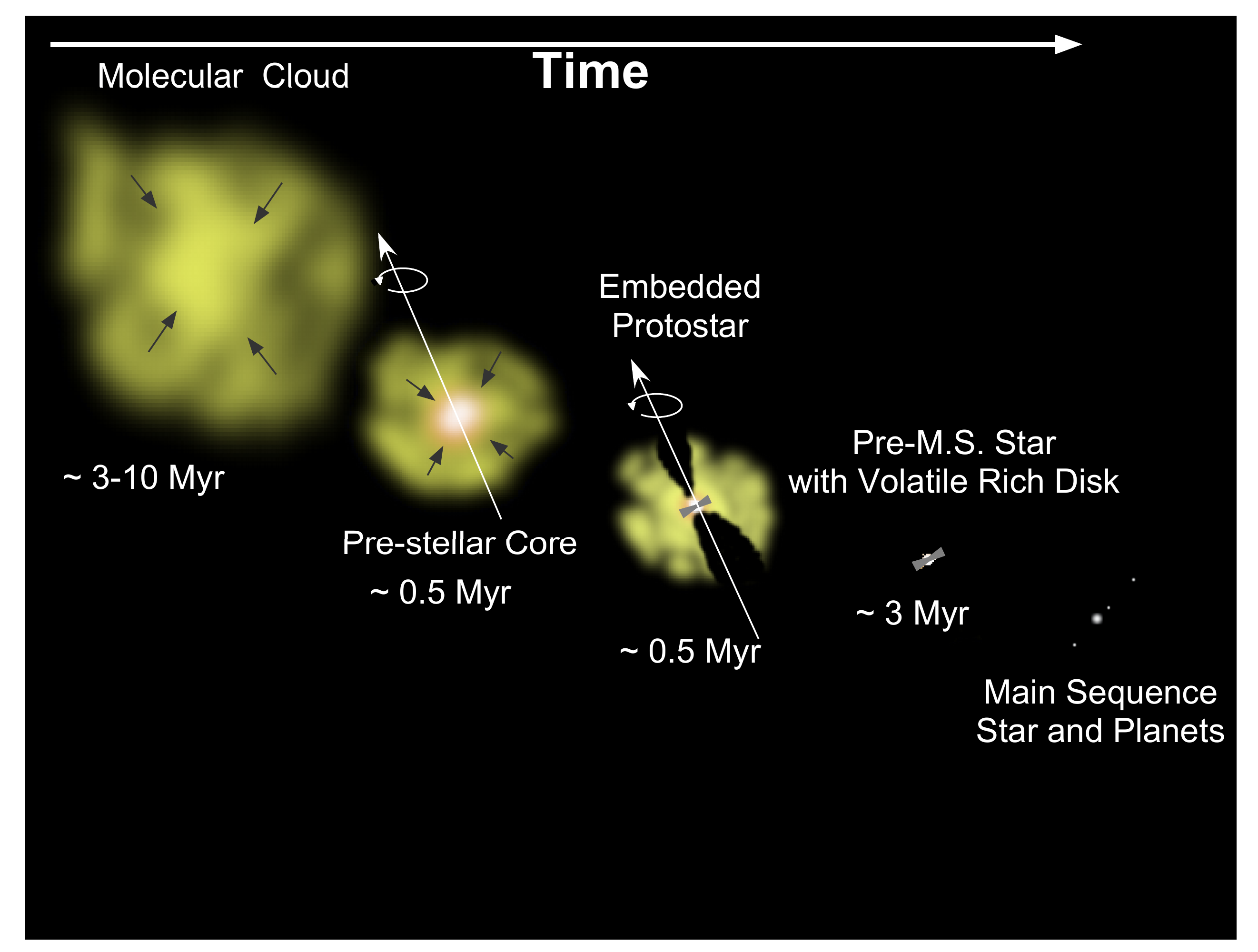}
\label{fig:schem}
\end{figure}

In our solar system, and in particular on the Earth, we can use a variety of techniques to look backwards in time and, where possible, sketch a chronological sequence.     
Astronomically, we can observe a large sample and measure the frequency of young stars and protoplanetary systems to construct an evolutionary sequence - in a way, looking forwards from beginning to end.

The interstellar medium, the material that exists between stars, is composed of mostly hydrogen gas, along with trace elements with an abundance commonly set by our understanding of the solar composition \citep{Asplund09}.
In the average galactic medium with a particle density of 1 cm$^{-3}$ the gas is traced in atomic form via the 21 cm spin flip transition of H~I and also by detection of heavy atoms via absorption lines in the ultraviolet \citep{Snow06}.
There is also abundant evidence of the presence of solid particulate matter with an average size of $\sim 0.1\mu$m called dust grains.    Dust in the interstellar medium is readily detected via  absorption of ultraviolet and visible starlight.  This energy is then re-radiated at far-infrared and submillimeter wavelengths.  For a summary of these issues see \citet{draine03}.

\begin{figure}
 \setcaptype{figure} 
\subfigure[Star Formation Phases]{
\includegraphics[scale=0.22]{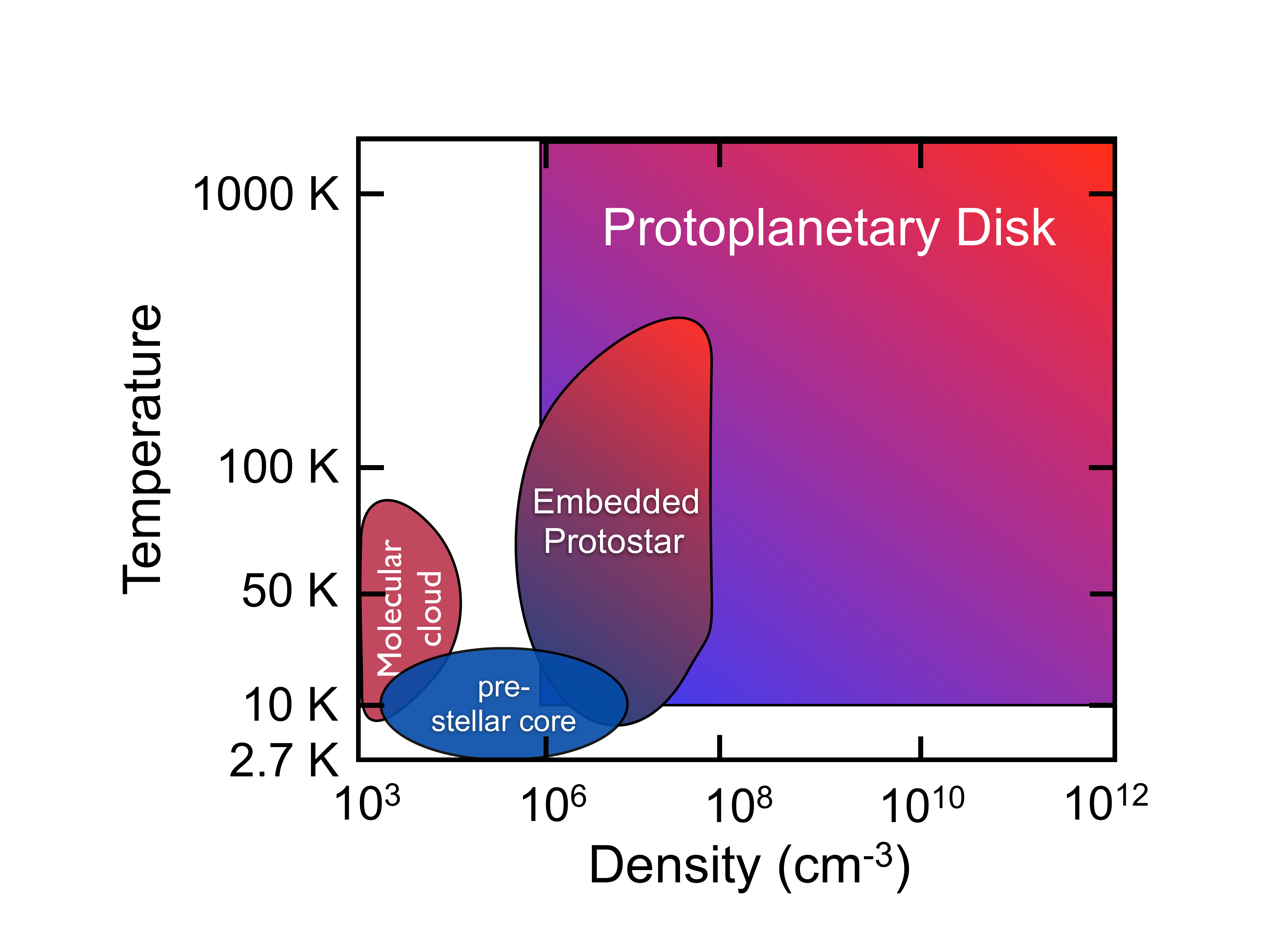}
\label{fig:specsub1}}
 \hspace{-0.45in}
\subfigure[Molecular Transitions as Probes]{
\includegraphics[scale=0.22]{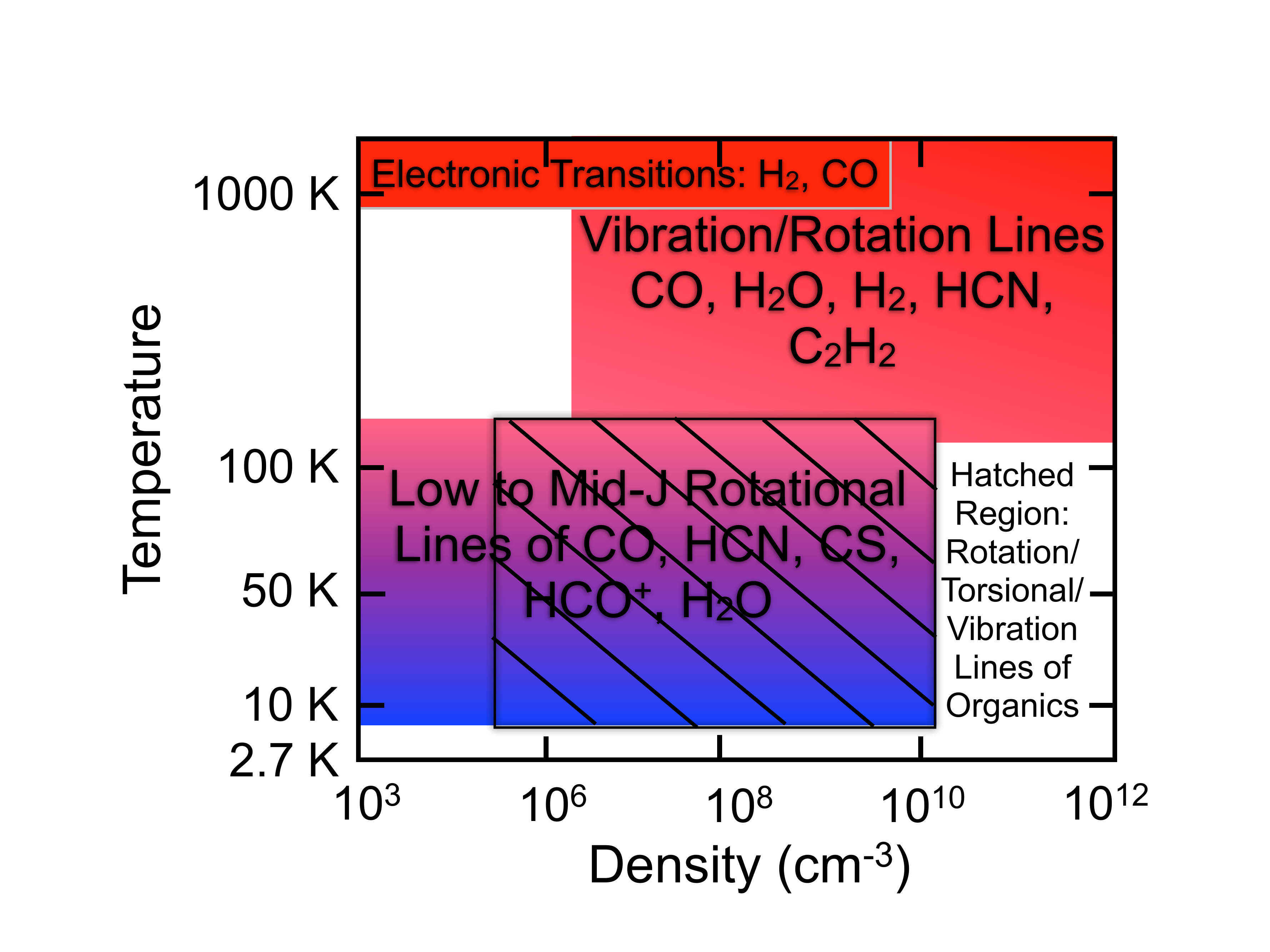}
\label{fig:specsub2}}
\caption{Plot illustrating the use of molecular transitions as probes of star and planet formation.  \subref{fig:specsub1} Rough outline of the physical conditions encompassed by the various 
phases in Fig.\ref{fig:schem}.   \subref{fig:specsub2}  Generalized physical regimes that can be probed by different types of molecular transitions when in emission.    
}
\label{fig:mp}
\end{figure}

In Fig.~\ref{fig:schem} we provide a schematic of the various phases and timescales in star and planet formation which are briefly described below.  A generalized visualization of the physical properties of each of these phases is given in Fig.~\ref{fig:mp}a.
Stars are born when portions of the very low density ($\langle n \rangle \sim 1$ cm$^{-3}$) warm (T $\sim 1000$~K) interstellar gas collapse or are gathered by some dynamical event that allows gravity to take hold \citep[e.g.][]{bp07, maclow04, Dobbs13}.
As part of this transition the gas becomes denser and shielded from the molecule-destroying interstellar radiation field, leading to molecular formation \citep{bergin_cform, clark12}.  We thus observe stars being born in what are called giant molecular clouds.    These complexes are vast covering many parsecs (1 pc = 3 $\times 10^{18}$~cm) with higher average densities ($<n> \sim 500-3000$ cm$^{-3}$) and lower temperatures (T $\sim 10-30$~K), due to the decreased penetration of starlight and the added cooling power of molecular radiation.    Lifetimes of giant molecular clouds are very much in debate with estimates ranging from a few to 20 Myr \citep[see][]{Dobbs13}.
 These clouds fragment and condense into successively small units with stars born in ``pre-stellar'' cores with characteristic sizes of $\sim 0.1-0.2$ pc.    These regions have higher central densities,  $n > $ few $\times 10^4$ cm$^{-3}$, but remain cold due to the lack of energy input into the center where the intensity of star light can be reduced by as much as 9 orders of magnitude.  
 Eventually gravity wins over support from thermal pressure, or the support from the magnetic field diffuses away, and the core subsequently collapse under the weight of their own gravity \citep{larson03, mo_araa}.   The evolutionary timescale estimated for this phase is $\sim 0.5$Myr \citep{Enoch08}.   

  Clouds are observed to be rotating from the large to small scale \citep{Arquilla86, Goodman93, Chen07} and collapse leads to the formation of a star and disk system \citep{tsc84} that is embedded inside its natal envelope.
 This phase is characterized by substantial thermal and density gradients with high values close to the young accreting protostar, decreasing with distance into the surrounding envelope.   The envelope is gradually dispersed over $\sim 0.5$ Myr \citep{evans09} leaving the star surrounded by an exposed volatile gas-rich protoplanetary disk.  The physical perspective for the disk phase is one with large variation in physical properties both radially (from the star and outward) and vertically (from the disk midplane and upward).   One key point is that the lifetime of the gas- or volatile-rich 
 phase has some uncertainty, but current estimates appears to be only a few Myr \citep{williams_araa}.\footnote{The birth environment of the Sun is sometimes called the Solar Nebula, which refers the the disk out of which planets are born.  We will refer to this phase as the solar nebular disk when talking specifically about the solar system and as the protoplanetary disk when discussing astronomical objects.}
  
 Molecules are present during each phase described in Fig.~\ref{fig:schem} and their emission can be used to probe the physics and chemistry throughout each stage of star and planetary birth.     The molecule interacts with the radiation and emits only if it has a permanent dipole moment which acts as a small antenna for radiating or receiving electromagnetic waves that correspond to its frequency of radiation.  
 The methodology for using molecular emission to trace the density, temperature, and velocity field of H$_2$, the dominant constituent is outlined in \citet{evans_araa}.    \citet{igh87} and  \citet{hvd09} outline how to determine the chemical composition of molecular gas.

  \subsubsection{The Physics of Molecular Spectral Lines}

 Molecular spectra are significantly more complex when compared to atoms.   The Schr\"{o}dinger equation is therefore also complex involving the positions and moments of all nuclei and electrons.   Thus molecules have rotational and vibrational motions along with electronic states that are a focus for atomic spectra.  
  Fig.~\ref{fig:mp}b graphically illustrates the regimes where various molecular transitions can be used to probe star formation.   Electronic transitions ($\Delta E \sim 1$~eV) are observed in the ultraviolet generally in absorption in diffuse low density molecular gas,  but also in emission in disk systems.    Vibrational modes have energies typically of order $\sim 0.01 - 0.1$~eV and can be observed in warm dense regions.   A vibrating molecule is also rotating and the emission occurs via a vibration-rotation cascade.    
  Low energy rotational modes of simple molecules are confined to the millimeter and submillimeter wavelengths, while higher energy rotational states trace higher temperatures/densities at shorter wavelengths.   One facet in this discussion is the rotational spacings, from low to high energy, are inversely dependent on the mass of the nucleus.  Thus lighter molecules such as H$_2$ or H$_2$O can have wide energy spacings between their transitions  (i.e. larger frequency)  while heavier molecules, HCN or CO, have closer energy differences between states.
  One caveat is that Fig.~\ref{fig:mp}b refers to regimes probed by transition of gas phase molecules. 
   Molecules are also observed in the solid state, i.e., as ices, by the same vibrational modes either in absorption using a star or cloud as a background candle \citep{gibb_iso, sonnentrucker08} and in a few instances in emission \citep{Malfait99, mcclure12}.

Molecules with a degree of symmetry have simpler spectra. 
Rotation can be described as motions about 3 principle axes, with 3 moments of inertia.   For many simple diatomic molecules, and linear polyatomic molecules,  the principle moment of inertia about the molecular bond axis is zero, and the moments of inertia about axes orthogonal to the bond are equivalent \citep{gordycook}.    In this case the rotational spectrum can be characterized by 
  rotational motions having spacings proportional to  $\Delta E = 2 B h J$, where $B$ is the rotational constant that is inversely proportional to the moment of inertia  ($B =$ 57.635 GHz for CO). $J$ is the  quantum number of the angular momentum that represents the upper state in this equation.  In contrast, NH$_3$ has two moments of inertia that are equal,  and is called a symmetric top characterized by two quantum numbers for a given energy state \citep{Ho83}.    Water is an asymmetric top; none of its principle moments of inertia are zero and no two are equal.  For asymmetric tops, the rotational states are characterized by three quantum numbers, and the resulting spectra can be very complex.     Large organic molecules (by interstellar standards!) - such as CH$_3$OH - have torsional motions where, for example, the OH undergoes an internal rotation relative to the CH$_3$ along with lower energy vibrational modes \citep{Koehler40}.

    \subsubsection{Molecular Astrophysics}

The observation of molecular emission bears information on the density, temperature, and dynamics of the emitting gas.
In certain cases there is also information on the total number of molecules that are emitting per cm$^2$, this is 
called the column density (denoted as $N$).
To interpret the measurement of the emission line strength of a given molecule, we must employ our knowledge of the molecular physics of the transition between two energy states.

  For a given transition between two energy states, we can define an excitation temperature, $T_{ex}$ according to the Boltzmann equation:
  
  \begin{equation}
  \frac{n_u}{n_l} = \frac{g_u}{g_l} e^{-h\nu_{ul}/kT_{ex}}.
  \end{equation}    
  
  \noindent Where $n_u$ and $n_l$ is the number density of molecules that exist in the upper and lower states between a transition with frequency $\nu_{ul}$; $g_u$ and $g_l$ are the statistical weights.   If the gas is in local thermodynamic equilibrium (LTE) then $T_{ex} = T_{k}$ (the gas kinetic temperature).   
   The intensity of radiation (erg s$^{-1}$ cm$^{-2}$ sr$^{-1}$ Hz$^{-1}$)  can be expressed by the common formalism for an isothermal medium (see nice discussion in \citet{kwok07}):
   
   \begin{equation}
I_\nu = I_{bg} e^{-\tau_\nu} + S_\nu (1 - e^{-\tau_\nu})
   \end{equation}
   
\noindent
$S_{\nu}$ is the source function which in this case is a blackbody, $B_\nu (T_{ex})$.  For long wavelength molecular emission (cm and mm-wave) the Rayleigh-Jeans approximation ($h\nu \ll kT$) can be used,
 In addition,  it is common  to measure on and off (background - $bg$) positions, taking a difference defining the on source position by a measured brightness temperature, $T_B$ = $\lambda^2 S_\nu$/2k:

\begin{equation}
T_{B} - T_{bg} = (T_{ex} - T_{bg}) [1 - e^{-\tau_\nu}].
\end{equation}

\noindent If the emission is optically thick then  the measured $T_B = T_{ex}$.  In LTE this is equal to the kinetic temperature.  We note that the Rayleigh-Jeans approximation does not always apply and the more explicit Planck function should be used for greater reliability.  Lets now explore the limit of low optical depth in the case where $T_{ex} \gg T_{bg}$ \citep[e.g.,][]{gl99},

\begin{equation}
\Delta I_\nu = T_B = B_\nu (T_{ex})\tau_\nu = \frac{h\nu}{4\pi} N_u A_{ul} \phi_\nu,
\end{equation}

\noindent where $\phi_\nu$ is the line profile function.  Integrating over the line we find, 

\begin{equation}
\int T_B d\nu  = \frac{h\nu}{4\pi} N_u A_{ul}.
\end{equation}

\noindent Emission is generally measured in velocity units instead of frequency, thus,

\begin{equation}
\int T_B d{\rm v}  = \frac{ hc^3}{8\pi k\nu^3} N_u A_{ul}.
\end{equation}

In summary, we have a relation between the upper state column density, $N_u$ and the measured quantity from the telescope which depends only on parameters set by the molecular physics: the spontaneous emission coefficient, $A_{ul}$ and the frequency, $\nu_{ul}$.  To convert this to the total molecular column, $N_{tot}$,  we must correct for the fact that there is a distribution of molecules that exist in various rotational (vibrational, ...) states and we have generally measured only one, or a small fraction.   For this we can use an expression that relates the fractional population, $f_u = N_u/N_{tot}$, to the total distribution:

\begin{equation}
f_u = g_u\frac{e^{-E_u/kT}}{Q(T)}.
\end{equation}

\noindent The partition function,

\begin{equation}
Q(T) = \displaystyle\sum\limits_{i} g_i e^{-E_i/kT},
\end{equation}

\noindent is summed over all $i$ energy states.   For simple and complex molecules and in LTE there exist approximations in the high-temperature limit for the partition function \citep{blake87, BTurner91}, or more explicit calculations are available in the catalogs by  \citet{pickett98} and, by \citet{muller05}.   

    \subsubsection{Determining Molecular Abundances}

   One of the interesting facets of molecular studies is that the primary constituent of molecular clouds, H$_2$, does not emit  at the cold temperatures of the gas ($\sim 10$~K).   This is because lowest rotational transitions has an energy spacing of $\sim 500$~K leading to negligible population in its first excited state and undetectable levels of emission.  The emission intensity is also reduced because H$_2$ has no permanent dipole, emitting instead via weaker electric quadrupole transitions.  Thus trace constituents are used as probes of the mass, density, temperature, and dynamics of the molecular hydrogen gas.   
   
   Given the central role of gravity in star formation, a key quantity in astronomy  is the overall mass. Furthermore in an astrobiological or astrochemical context we would like to trace the chemical abundance of a specific molecular species relative to the main gas constituent, H$_2$.
          Thus we need to have methods to approximate the amount of H$_2$ that is present.   A variety of techniques are used including using the emission and absorption of dust particles, to be discussed below.
   Here we discuss another common method, 
    the use of another widely abundant gas phase molecule, such as CO, as a tracer of H$_2$. 
    
To estimate molecular abundances the column density of molecular hydrogen needs to be estimated.   One method has been to calibrate the abundance of optically thin isotopologues of carbon monoxide\footnote{The emission of $^{12}$CO is typically optically thick.} to that of dust which has a separate calibration to H.   In this fashion the abundance of C$^{18}$O is estimated to be $\sim 1.7 \times 10^{-7}$ relative to H$_2$ \citep{flw82}.  Assuming an isotopic ratio of 500 the $^{12}$CO abundance is $\sim 10^{-4}$, which is measured along numerous lines of sight \citep{ripple13}.     Thus with observations of optically thin CO isotopologue emission and any molecular species, we can obtain abundance estimates: $N_{\rm X}/N_{\rm H_2}$.    More sophisticated techniques are now being adopted (see \citet{bt_araa}) but this offers an outline of the methodology.

The derivation of mass then only requires maps to be made of a probe that has an abundance calibrated to H$_2$.  Thus for example maps of $^{13}$CO emission provide the distribution of $N_u(x,y)$.  Using $^{12}$CO emission to estimate temperature, LTE, and a CO abundance, this can be converted to $N_{\rm H_2}(x,y)$.  The mass is then $M = 
\mu \displaystyle\sum\limits_{x,y} N_{\rm H_2}(x,y)\pi r^2$, where $r$ is the radius of the antenna beam corrected for the source distance, and $\mu$ is the correction for He and trace elements to the mass.

  \subsubsection{Properties of Interstellar Dust}

 Interstellar dust grains become incorporated into terrestrial planets, but also play a major role throughout star formation by mitigating the effects of energetic starlight and providing a surface to facilitate catalytic chemical reactions.   The emission of dust is also a powerful supplemental method to trace evolution and structure in star/planet formation.
 Dust grains also lock up key portions of the interstellar carbon and oxygen budget into refractory molecules.     For background we provide a brief discussion of the mitigation of starlight by dust grains along with the overall budget that is used to provide context for our discussion of the creation of molecular species in space.

   Dust absorption of starlight is treated in terms 
of magnitudes of extinction, labelled as $A_{\lambda}$ in units of magnitudes.  We define magnitudes such that a difference of 5 mag between two objects corresponds to a a factor of 100 in the ratio of the fluxes.   Thus an extinction of 1 mag corresponds to a flux decrease of $(100)^{1/5}  \sim 2.512$.  If we take a cloud exposed to the interstellar radiation field \citep{mathis_isrf, Draine_ism} characterized by an intensity,  $I_{\nu, 0}$, in units of ergs s$^{-1}$ cm$^{-2}$ \AA$^{-1}$ sr$^{-1}$.
As light propagates into a molecular cloud it is attenuated by dust absorption, $I_{\nu} = I_{\nu,0}{\rm exp}(-\tau_{\lambda})$.   The opacity,  $\tau_{\lambda}$, is given by 
$\tau_{\lambda} = N_{dust} Q_{\lambda} \sigma $.   Here $\sigma$ is the geometrical cross-section of a single grain, 
$Q_{\lambda}$ is the extinction efficiency, and $N_{dust}$ is total dust column along the line of sight.
Now lets put  $A_\lambda$ in terms of optical depth:

\begin{eqnarray}
A_{\lambda} & =& -2.5{\rm log}(I_{\nu}/I_{\nu, 0}) \\
 &=& 2.5 {\rm log}(e) \tau_{\lambda} \\
  &=& 1.086 N_{dust} Q_{\lambda} \sigma \\
  &=& 1.086 \frac{n_{dust}}{n} N Q_{\lambda} \sigma. \label{eq:al}
\end{eqnarray}

\noindent  In this equation $n_{dust}$ is the space density of dust grains, $n$ the space density of the gas, and $N$ is the total gas column density.    We can also convert $N_{dust}$ to $N$ via a dust-to-gas mass ratio, $\xi$.
This is estimated to be $\sim 1$\% in the ISM \citep{Savage79} (and can vary inside a protoplanetary disk) \citep{Furlan05}.

\begin{equation}
\frac{\rho_{dust}}{\rho_{gas}} = \xi = \frac{n_{d}m_{gr}}{n \mu m_H},\\
\end{equation}

\begin{equation}
\frac{n_{d}}{n} = \frac{3}{4\pi} \frac{\xi \mu m_H}{\rho_{gr} a_{gr}^3}.
\end{equation}

\noindent Rearranging Eqn.\ref{eq:al} and taking typical values where $Q_V \sim 1$ and a grain size of 0.1 $\mu$m, and $\rho_{gr} = 2$ g cm$^{-3}$ then,

\begin{equation}
A_V = 1.086  \frac{3}{4} \frac{\xi \mu m_H}{\rho_{gr} a_{gr}} N Q_{V} \sim \left(\frac{\xi}{0.01}\right)\left(\frac{Q_V}{1}\right) 10^{-21}N.\label{eq:taud}
\end{equation}

 \noindent Thus 1 mag of extinction at visible wavelengths is provided by a total gas column density of $10^{21}$ cm$^{-2}$.    This is consistent with observations of dust extinction and hydrogen absorption \citep{Bohlin78}. 
Dust absorption is wavelength dependent.  In Eqn.~\ref{eq:al} this is codified in the wavelength dependent extinction efficiency.  Extinction of starlight is greater at shorter wavelengths (compared to visible) and reduced at longer wavelengths.   
The above relations are also simplified as the  grains are inferred to exist with a distribution of  sizes following a power-law of $dn(a)/da \propto a^{-3.5}$, with $a$ as the grain radius \citep{mrn}.    Thus detailed models of the composition and size distribution of grains are used \citep{weingartner_draine01}.
  At sub-millimeter wavelengths the dust emission is optically thin and can be used to probe the gas density structure and total mass, assuming a gas-to-dust mass ratio \citep[e.g.][]{bt_araa}.  
 
The composition of interstellar dust grains is estimated via observations of low extinction interstellar gas clouds that are seen in absorption via various atomic transitions towards background stars.  Since \ion{H}{I} and H$_2$ can be observed directly in this gas via electronic absorption lines the abundance of a given element can be estimated and compared to our standard, the abundance of elements in the Sun \citep{Asplund09}.   These observations have shown that abundances of refractory materials (Mg, Si, Fe) are severely depleted from the gas, and presumably reside in dust grains \citep{ss96}.   These depletions also correlate with the condensation temperature of the given element, which is similar to that found from analysis of the composition of meteoritic rocks \citep{Ebel00, lodders03, Davis_geochem}.    More volatile material exhibits less gas phase depletion, but a large portion of O and C are locked in refractory materials: O in the form of silicates and C either in amorphous carbon or polycyclic aromatic hydrocarbons \citep{draine03}.
The abundance of carbon, oxygen, and nitrogen measured in the Sun is, respectively, 245, 537, and 72 parts per million (ppm, i.e. relative to 10$^6$ H atoms) \citep{Lodders10}.    
\citet{whittet10} estimates that $\sim 170$ ppm of the carbon and $\sim$ 140 ppm of the oxygen is locked up in these refractory carbonaceous and  silicate materials (respectively).   N and H remain in the gas.
    
  \subsection{The Interstellar March Towards Complexity}

Below we will explore each of the phases of star formation into the protoplanetary disk phase discussing the overall understanding and disposition of the key molecules that probe the CHON pools.
 
  \subsubsection{Molecular Cloud}
The distribution of mass and the overall reservoir available for stellar birth in the giant molecular cloud is traced primarily by the emission of carbon monoxide.     
     In our galaxy one generally obtains emission maps of $^{12}$CO and its isotopologue $^{13}$CO.    The emission of carbon monoxide is optically thick, tracing temperature instead of gas mass\footnote{On very large scales when observations are not resolved, capturing all gas motions, $^{12}$CO emission can trace the mass though the linewidth and surface area (see the nice discussion by \citet[][]{Draine_ism}).  This has been widely used to trace the star-forming gas mass in other galaxies \citep{kennicutt_evans}. }.     Based on this we find that molecular clouds are massive, containing 10$^{4}$ - 10$^{5}$ M$_{\odot}$, enough mass to form clusters of stars \citep{Blitz99, 2003ARA&A..41...57L}.    There exists clear hierarchical structure with the densest portions occupying less volume then lower density material.

     \begin{figure}
\caption{Spitzer's infrared view of the sword in Orion: the Orion Nebula.  In the image the red wispy material is the dust grains heated by starlight -- illustrating the complex structures in star-forming regions.   Dark lanes superposed on this emission represent the dense regions of the Orion Molecular Cloud where stars are currently being born (protostars).
Image credit: NASA/JPL-Caltech/T. Megeath (Univ. of Toledo) but see also \citet{Megeath12}.}
\includegraphics[height=0.6\textheight]{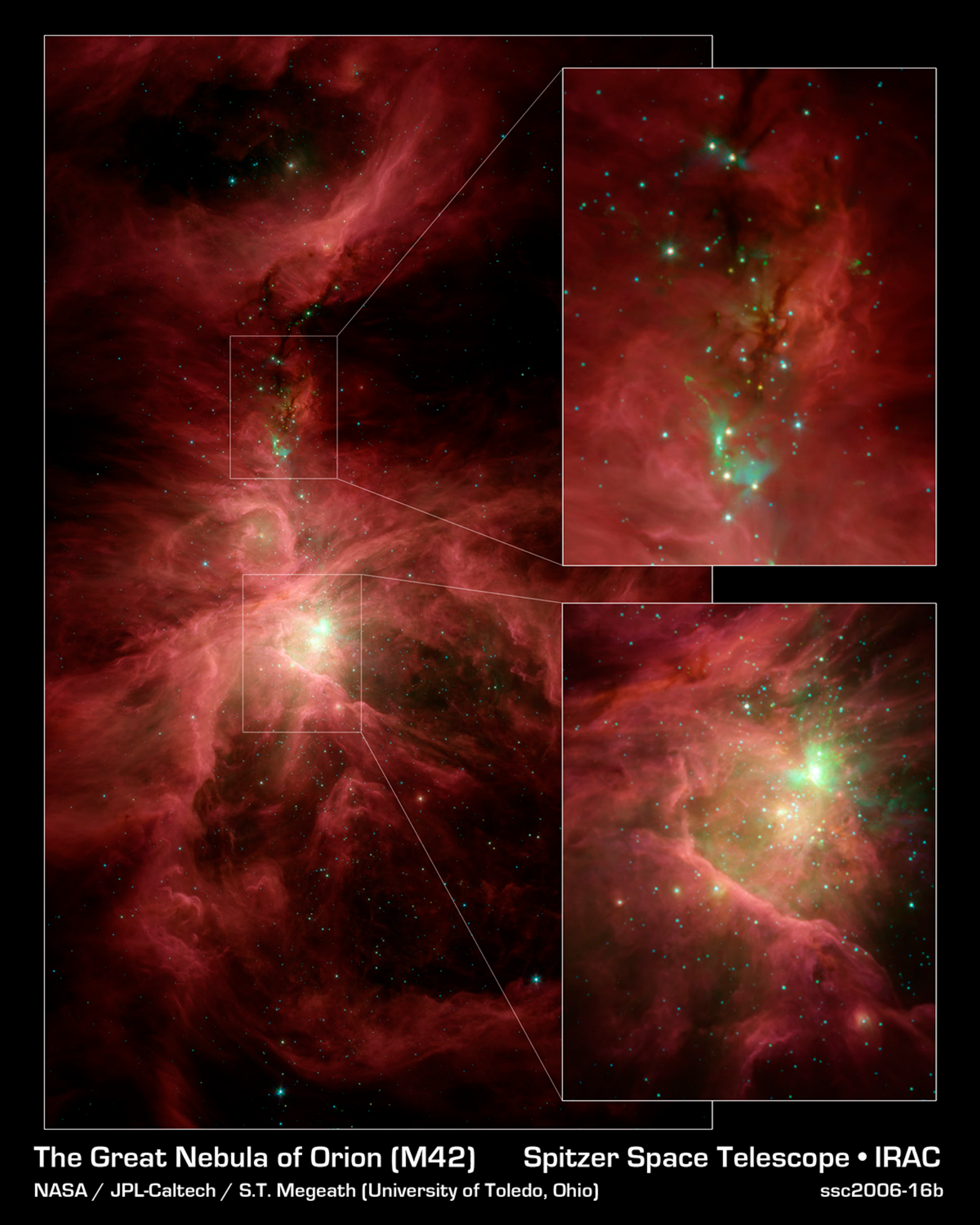}
\label{fig:orion}
\end{figure}

\begin{figure}
\vspace{-1in}
\caption{A deep optical image of the dark globule
Barnard 68 ({\em top left}; \citet{alves_b68}) along
with contour maps of integrated intensity from molecular emission 
lines of N$_2$H$^+$, C$^{18}$O \citep{bergin_b68}, and 850$\mu$m dust continuum emission \citep{bianchi_td}. The angular resolution of the molecular data is $\sim 25''$ and the dust emission is 14.5$''$.  
Figure first published in \citet{bt_araa} and we reproduce that caption here.}
\includegraphics[height=0.6\textheight]{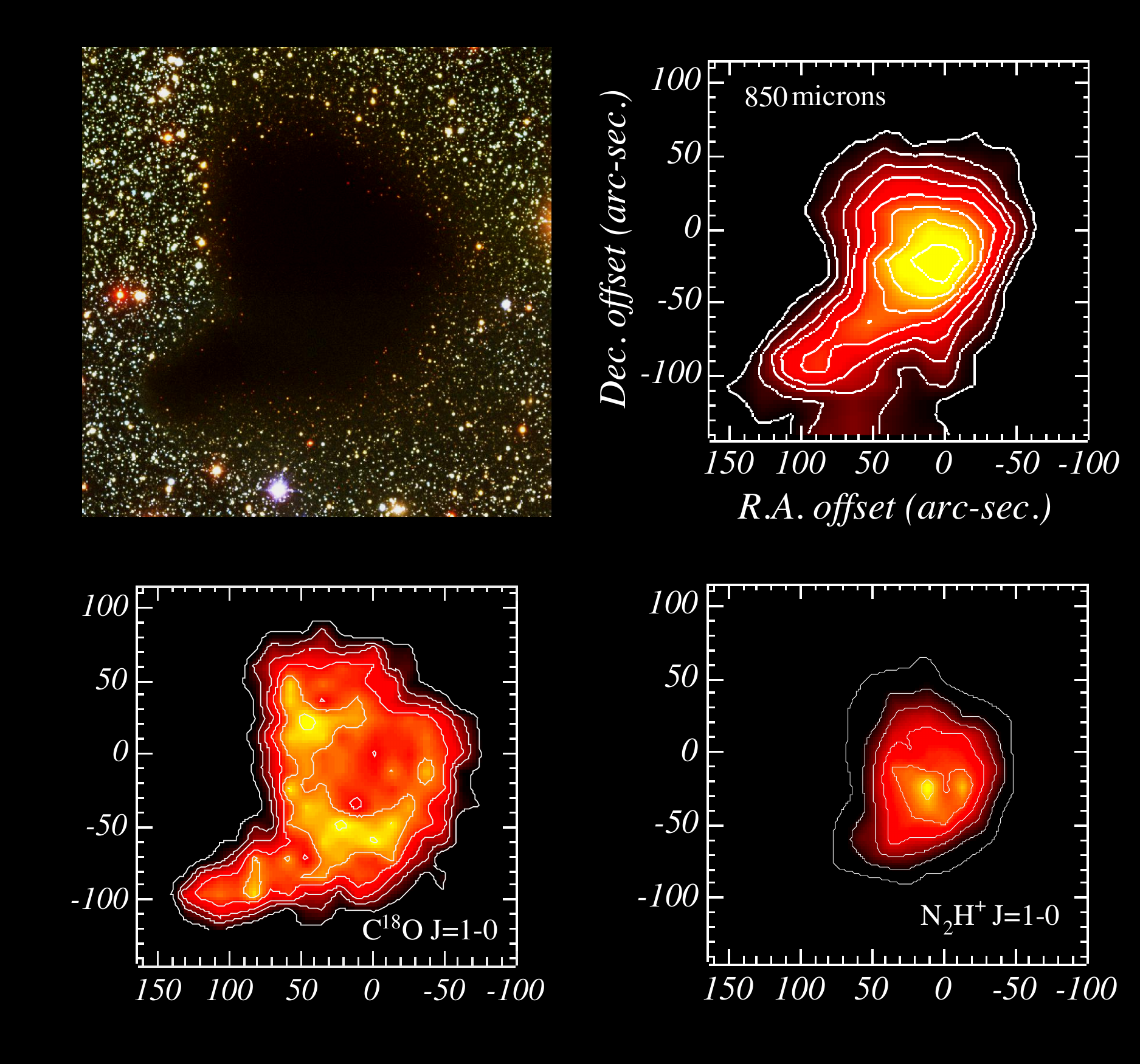}
\label{fig:b68}
\end{figure}

 Infrared studies of young embedded stars provide the direct association of the densest portions of this molecular material with stellar birth \citep{Lada92}.    
   Fig.~\ref{fig:orion} shows one wide field survey of the Orion molecular cloud.  This striking image illustrates the complexity of structures in star formation as the red material in this image is the dust heated by starlight.   The central region is dominated by the Orion nebula where thousands of stars formed in the last million years.      Towards the north we observe dark lanes superposed on the background emission from dust.  This is the region where stars are being born (see bright sources in the inset to the top right).  If we observed this region in molecular emission it would shine brightly, e.g. Fig.~\ref{fig:b68}.
   Such an environment provides important context for the formation of our own solar system, as it is thought that the Sun was originally a member of a massive stellar cluster \citep{2010ARA&A..48...47A}.
 
 In this phase nearly all volatile molecules not locked into the refractory component are found in the gas.  
 Gaseous molecules form in the general low density ($n \sim 1000 - 3000$ cm$^{-3}$) cloud material via two mechanisms.  
 In the case of H$_2$ a search for viable gas phase pathways finds no mechanism capable of creating H$_2$ from two H atoms in a requisite time frame.  Thus H$_2$ is believed to form  through catalytic reactions on the surfaces of the cold (T $\sim 10-20$~K) dust grains \citep{hs71}.  The viability of this mechanism has been confirmed by laboratory experiments \citep{Katz99}. 
  For many other simple molecular species, the main formation pathways are thought to occur via two body ion-neutral reactions in the gas phase \citep{hk73}.    These reactions are fast and interstellar cosmic 
rays provide a source of ionization.   Reaction networks based on these theories have been constructed and follow the transfer of, for example, C$^+$ into C and then CO via series of reactions \citep[e.g.][]{Herbst95}.  

A key facet in the question of molecular formation is the destruction process.   In regions exposed to the interstellar ultraviolet radiation field the process of photodissociation dominates and for the most part atoms and ions exist.  For a given species its ionization state will depend on whether its ionization potential is above or below the Lyman limit (13.6~eV).  Here the main differences are whether a molecule is dissociated by a line process or by a continuum of photons \citep{vd87}.   In the former case, with sufficient column molecular lines can become optically thick.  Thus molecules closer to the surface can shield molecules deeper in the cloud from the effects of destructive radiation.     The most abundant gaseous molecules, H$_2$ and CO,  are both dissociated though lines and exist in regions where the dust optical depth to UV photons is low \citep{Lee96, visser09}.   Molecules dissociated via a continuum process (e.g. H$_2$O, NH$_3$, organics) require the shielding by dust grains  (Eqn.~\ref{eq:taud}).
 Deep inside clouds the optical depth due to dust grains is large ($A_V > 3$), and the effects of external radiation are negligible; in this case  molecules are typically destroyed via reactions with He$^+$ \citep{Langer89}.

    \subsubsection{Pre-stellar Core} These dense molecule-dominated condensations that have clear evidence for a high degree of central concentration in the gas and dust density.    
    Fig.~\ref{fig:b68} shows a template object for this class: Barnard 68.    
 In the optical image (top left) we observe that the core is optically thick as the rich field of background stars cannot be observed towards the center where $A_V > 10$ mag \citep{alves_b68}.    The dust grains that absorb starlight conserve energy and thus  warm up, remitting this emission in the far-infrared/submillimeter.  The top right panel shows the submillimeter continuum image of the reprocessed starlight.     As noted above both the emission and absorption trace the total dust column and thus $N$ (assuming a dust-to-gas mass ratio).   Since the emission is centrally concentrated the column density, and hence volume density, is also higher towards the center.  Estimates suggest volume density increases of nearly two-three orders of magnitude from edge to center \citep{difran_ppv, bt_araa}.  
Studies of the dust temperature (by fitting a black body) and gas temperature (molecular lines) demonstrate that these objects are very cold with gas and dust temperatures in their centers of $\sim 10$~K \citep{bergin_tgas, Crapsi07, Launhardt13}.
 
An example of the changing chemistry of these objects is shown in the bottom two panels.   The emission distribution of C$^{18}$O has a minimum towards the core center, while N$_2$H$^+$ (a molecular ion that is a chemical daughter product of N$_2$) traces the core center more directly, albeit with structure.    These changing distributions are the result of the interaction of gas-phase molecules with cold dust grains.  In particular the timescale for a molecule to freeze out onto the surfaces of cold dust grains becomes significantly shorter than the cloud lifetime.     For an interstellar grain size distribution \citep{mrn}, $n_d \sigma \simeq 2 \times 10^{-21}n$ cm$^{-1}$ \citep[see discussion in][]{hkbm09}.     Assuming this surface area per H, the freeze-out timescale, $\tau_{\,fo}$, is:
 
\begin{equation} 
\tau_{\,fo} = 5 \times 10^3 {\rm yrs} \left(\frac{10^5 {\rm cm}^{-3}}{n}\right) \left(\frac{20 K}{T}\right)^{\frac{1}{2}}\sqrt{A}
\end{equation}

\noindent where $A$ is the molecule mass in atomic mass units.    Thus in the center of pre-stellar cores there is a 2-3 order of magnitude decrease in the freeze-out timescale.  The hole in the C$^{18}$O distribution is essentially due to CO beginning to preferentially freeze-out onto dust grains in the denser central regions of the core \citep{bergin_b68}.   This hints at a key part of the chemistry of this phase: formation of ice coated grain mantles, as the evidence shown above is now known to be widespread \citep{tafalla_dep, bt_araa}.    Water, CO, and CO$_2$ ices are also detected in absorption towards bright background sources  and may begin to form in the cloud itself \citep{whittet98, klaus04}.
Water ice is the dominant molecule in the ice mantle, similar to that seen for cometary ices.    Water vapor is now known to be a minor constituent of the gas in pre-stellar cores \citep{Caselli12}.    Based on these observations and models, it is during  this phase when water is made that becomes seeded to the disk as ice upon collapse.      Some ices, such as water, form via a complex set of reactions on grain surfaces, while others such as CO form in the gas and then freeze onto grain surfaces \citep{Cuppen10, Cuppen11}.   However, once on the grain CO can also participate in surface chemistry ultimately leading to methanol and other organics \citep{hvd09, Oberg11c}.   

At the cold, $\sim 10$~K, gas temperatures present in the core ion molecule reactions lead to high levels of deuterium enrichments.   Deuterium fractionation is driven by the following reaction:

\begin{equation} \label{eq:h2dpform}
\rm{H}_3^+ + HD \leftrightarrow  \rm{H}_2D^+ + H_2 + 230 K.
\end{equation}
 
 \noindent  The forward reaction  is slightly exothermic favoring the production of H$_2$D$^+$ at 10 K, enriching the [D]/[H] ratio in the species that lie at the heart of  ion-molecule chemistry \citep{millar_dfrac}.    CO rapidly destroys H$_3^+$, inhibiting the formation of H$_2$D$^+$.    Hence, formation of CO ice has been found to be important in increasing the rate of the sequence of gas-phase fractionation reactions  \citep{roberts_dfrac}.   Thus high level of enrichments in a variety of species are found in this phase \citep{Roueff03}.  As one example, DCN/HCN $\sim 0.01$ in several cold cores, an enrichment of 3 orders of magnitude.   In addition, a by-product of the gas phase process, is an enhancement of D atoms relative to H atoms that are then placed into ices as they form on the cold grains via catalytic reactions.  Thus any ices forming via grain surface chemistry will have high D/H ratios 
  \citep{ta87}.

{\em In conclusion, the beginnings of chemical complexity potentially required for astrobiology are linked to the earliest phases: the formation of  water and initial organics -- with deuterium enrichments seen in solar system ices.}

      \subsubsection{Embedded Protostar} 
   The delivery of volatiles, predominantly in the form of ice, from the pre-stellar core to the protoplanetary
disk occurs during the embedded protostellar phase. This phase begins with the gravitational collapse of
a molecular core and the formation of an accreting protostar in the
center of the core, surrounded by a disk \citep{als87, awb93}.   Protostars can be readily detected via the strong continuum emission as the accretion luminosity is absorbed by the dense ($n > 10^5$ cm$^{-3}$) envelope of gas surrounding the star (e.g. Fig.~\ref{fig:orion}), reprocessed by  the dust and reradiated in the mid-infrared to submillimeter  wavelengths \citep[$20 - 1000\;\mu$m;][]{evans09, Bontemps10, 2010A&A...518L.122F, Maury11, Megeath12}.

   \begin{figure}
\caption{{\em Top panel:} Herschel/HIFI full band spectrum of Orion KL to be published in Crockett et al. 2013 (in prep.).  {\em Middle panel:}  zoom into one area of the spectrum with a handful of molecular species labelled with associated transitions.  This plot has been published earlier in \citet{bergin10a}.
{\em Bottom panel:}  Final zoom into the region of the spectrum coincident with a Q-branch band of CH$_3$OH.   This spectrum was presented and analyzed by \citet{wang11}.}
\includegraphics[height=0.9\textheight]{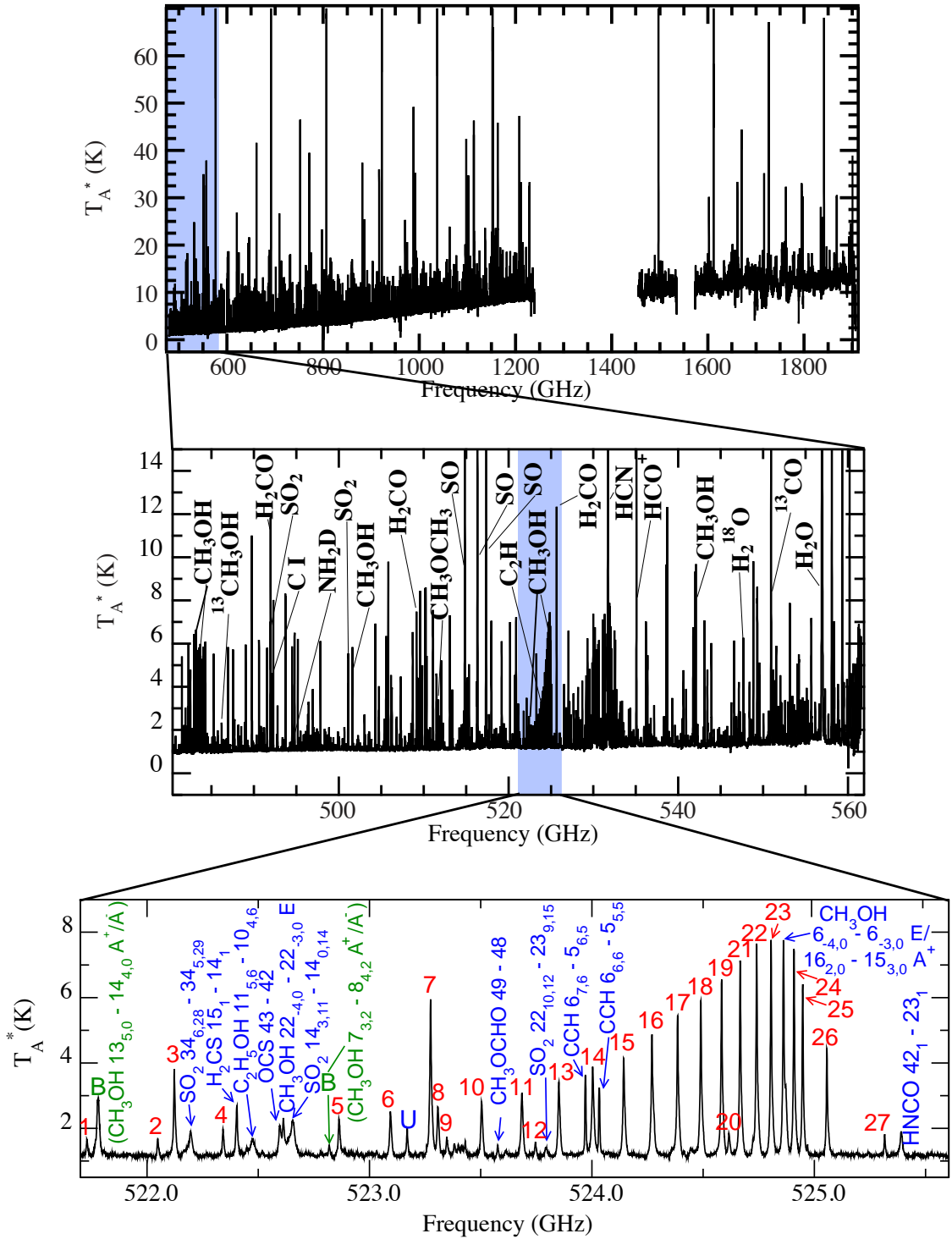}
\label{fig:kl}
\end{figure}

   The heating from a luminous central source  leads to strong thermal gradients in addition to the high degree of central concentration in the density.   These conditions are ripe to excite the emissions of numerous molecules and it is during this phase that we gain the greatest glimpse of the chemical complexity that is fostered in interstellar space.  
In fact when the temperatures rise to $> 50 - 100$~K the contents of the icy mantles coating grains are released. 
It is in these regions, so-called ``hot cores'', where most molecular species have been detected in space.   At present over 170 molecules have been detected, with a clear dominance of organics.

 This richness of interstellar chemistry is prominently observed towards young protostars more massive than our Sun \citep{blake87}, but  also toward low mass regions \citep{hvd09, cc12}.   Fig.\ref{fig:kl} shows one such example in the Herschel spectrum \citep[][Crockett et al. 2013, in prep.]{bergin10a} taken toward a massive star-forming region found in the center of the image in Fig.~\ref{fig:orion}.   In this one spectrum alone 20,000 lines are observed arising from 35 molecules (84 isotopologues).   The richness of the spectra in these regions with overlapping lines and emission down to the sensitivity limits presents tremendous challenges in analysis.  Based on observations like these we now know that interstellar chemistry  extends to species of astrobiological interest.  This includes amino-acetonitrile,  a pre-cursor to glycine, the simplest amino acid \citep{Belloche08}; glycholaldehyde, a simple sugar, has been detected from gas in close proximity to a solar mass type star and also towards more massive objects \citep{Hollis00, Jorgensen12}.   In addition E-cyanomethanimine, an HCN dimer, has been also be observed \citep{Zaleski13}.  This species is a key intermediary in potential formation routes to adenine, a nucleobase.   

Both evaporated water and organics exhibit high levels of deuterium enrichments as the ISM values shown in Fig.~\ref{fig:dh} are predominantly from detections in this stage.   The main deuterium fractionation pathways, originating via H$_2$D$^+$ (and also CH$_2$D$^+$), require temperatures less than $\sim$50~K to be operative.  This is inconsistent with the temperature inferred from the emission of both water and organics (see references in Fig~\ref{fig:dh}) which are generally $> 100$~K.    Thus these deuterium enrichments are believed to be reflective of chemistry that has occurred during the earlier colder pre-stellar phase.  For organics the level of enrichment seen in the ISM is significantly higher than that seen in the meteoritic matrix.    For water, measurements of this ratio towards young solar type protostars have found values significantly closer to that of  Oort cloud comets (e.g \citet{Persson13} but see also \citet{Taquet13}).

 Based on these observations it is theorized that many organics are created via surface
chemistry in the pre-stellar phase, along with water.  Over the subsequent several hundred thousand  year evolution bonds within the ices can be broken via photoabsorption\footnote{Even in dark regions of space cosmic rays generate a weak UV radiation field \citep{Prasad83}}, freeing radicals which cannot move freely on grain surfaces at 10 K.  Once the star is born the grain warms up, heavier radicals gain needed energy to move on the grain surface opening up additional organic creation pathways \citep{gwh08, hvd09}, leading to even greater chemical complexity.
We note also that there are some limits in our ability to detect very large organics via resolved spectroscopic techniques.  Large molecules (e.g. amino acids, DNA bases) have numerous decay pathways which spread the excitation over many routes leading to weaker lines that can be lost in the line forest of more abundant molecules or below our sensitivity limits.

      \subsubsection{Protoplanetary Disk}   Over a timescale of $\sim$0.5~Myr the surrounding gaseous envelope dissipates and the volatile-rich gaseous disk becomes exposed.   Typical sizes of disk systems are $\sim 100$ AU \citep{williams_araa} and thus, even in the closest star forming region (Taurus at 140 pc) they subtend small angles on the sky, which makes observational characterization challenging.  
      Nonetheless disks are readily detected by the excess of emission at infrared wavelengths that exceeds that expected for stellar radiation \citep{Strom75, als87}.    This excess emission is the UV and optical stellar light that is reprocessed by the circumstellar disk and reemitted in the infrared by the warm dust grains.
       
      For the first few Myr the star is still accreting material from the disk \citep{muz00}.  Due to stellar irradiation and the release of accretion energy  the gas-rich protoplanetary disk has strong radial and vertical thermal gradients.   \citet{kh87} showed that the thermal energy decreases more slowly with increasing radius than the vertical component of gravity.   As a consequence,  the disk flares slightly with increasing radius, a fact that is confirmed by Hubble Space Telescope images \citep{Burrows96, Padgett99}.
  Most of the mass resides in the midplane of the disk and the surface density of material ($\Sigma$ in g cm$^{-2}$) decays with distance.   In the solar system this is codified by the Minimum Mass Solar Nebula, the minimum amount of material needed to make our planetary system, which has $\Sigma \propto r^{-3/2}$  \citep{w77_mmsn, hayashi_mmsn}.    
  Observations of optically thin submillimeter-wave dust emission can be used to trace mass in protoplanetary disks and these disks exhibit similar dependencies in the mass distribution within the errors \citep{andrews09, williams_araa}.   
      
      There are a number of reviews of the physics and chemistry of these systems that are available \citep{cc_messii, bergin_ppcd, Semenov12} so we will not get into specifics, but rather highlight a few key points.  (1)  Our current observational census finds that some disks have radii $> 100$ AU \citep{williams_araa} and are thus larger than our solar system as defined by the outer boundary of the Kuiper belt at 50 AU \citep{abm02}.   (2)       The evolution of dust in terms of planet formation is a major factor in the physical and also chemical evolution of the system.  (3) The system is not static.   Dust grains, until they reach $\sim$km sizes, are subject to a variety of forces that produce both random and systemic motions \citep{cc06, Ciesla12}.  Similarly the gaseous disk will have varying degrees of turbulence \citep{Balbus11}, which can lead to mixing \citep{Willacy06, Semenov11}.   As we noted earlier there is also likelihood of strong movement of material once the gas disk dissipates, under the influence of giant planets.  (4) 
Because of the flared surface the disk intercepts greater amounts of stellar radiation than would otherwise be the case.  This is particularly important for the chemistry as the disk becomes exposed to energetic stellar ultraviolet (UV) and X-ray radiation.   During this phase of evolution the UV radiation is dominated by stellar Ly $\alpha$ photons \citep{herczeg_twhya1, bergin_lyalp, Schindhelm12}.
Laboratory experiments suggest that  the exposure of interstellar ices to ultraviolet radiation leads to the formation of monomers, e.g. amino acids \citep{Bernstein99, 2002Natur.416..401B}.  Furthermore, it is clear that this radiation exists and is hitting the molecular disk
 \citep{France12}.
(5) The presence of radial thermal gradients suggests that there may be several condensation fronts for major volatiles with water closer to the star and, for example, CO at greater distances \citep{Qi13_sci}.      
 This list could be made much longer and we can summarize by stating that there are ample opportunities in the disk system to alter the chemistry within the volatile pools.

 In terms of observations we do have some information.       In the outer parts of the disk (radii $>$ 50 AU) there are detections of simple molecules in numerous systems \citep{oberg_discs1, oberg_discs2, Guilloteau13} (e.g. HCN, HCO$^+$, H$_2$CO).  Detailed studies suggest that molecular abundances are reduced compared to that of the interstellar medium, implying that the formation of cometary ices has commenced in these $\sim$1 Myr old systems \citep{dutrey_ppv, bergin_ppv}. 
  The characterization of water vapor has been a success story as we expect H$_2$O vapor inside the snow-line and ice beyond.  This is consistent with observational results  of   warm ($\sim 300- 500$~K) water vapor emission, arising from within a few AU of the star \citep{cn08, salyk08, pascucci09, pont10, salyk11}; at large radii water appears to be mostly frozen onto dust grains \citep{bergin10b, hoger11a, zhang13}.    These water observations are unresolved, so the resolved observation of the CO snow-line in one system \citep{Qi13_sci} using  the newly minted Atacama Large Millimeter Array (ALMA) points to a bright future of resolved chemical studies. 
  Another clear success is that resolved observations of molecular emission in the disk very beautifully trace the velocity structure and set limits on the physics of the disk and even constrain the stellar mass \citep{sdg00, rosenfeld12, Casassus13}.

   \subsection{The Chemical Legacy from Star and Planet Birth}
 
As seen from above there are some areas where we have obtained strong constraints on the creation and distribution of
volatiles, although during some phases we lack observational constraints.    However,  the interstellar record does show the
beginnings of chemical complexity was present and fostered as part of the stellar and planetary birth, which we summarize below through a census the major atomic pools.

\begin{itemize}

 \item {\it Hydrogen}: Formed as part of the birth of the molecular cloud and resides in the gaseous state in the form of H$_2$ throughout all stages.  \\

\item  {\it Carbon}:  A large fraction of the carbon is found in the solid state in molecular clouds, with estimates
as high as $\sim 69$\% in some form of solid state carbonaceous grain material.  
In regions exposed to starlight  broad spectral features are observed between 2-20$\mu$m.
 These have  generally been  associated with polycyclic aromatic hydrocarbons (PAHs), which are
 widespread in the ISM \citep{Boulanger88}.  We note that the red wispy dust emission in Fig.~\ref{fig:orion} is attributed mostly to emission from PAHs.   
 Models and observations suggest about 20-30\% of the carbon that is locked in the solid state resides in this unidentified aromatic form \citep{Habart04}.   The remainder of the carbon reservoir is found in gas phase carbon monoxide ($\sim 31$\%), which sets the stage for the birth of the dense core.
 
 Inside the dense ($n > 10^4$ cm$^{-3}$) cold (T $\sim 10-20$~K) pre-stellar core CO freezes onto grains and a fraction of this carbon is transferred into organics predominantly via catalytic chemistry on grain surfaces.    PAH's may also freeze onto larger grains in this stage \citep{Bouwman10}. These organics incorporate the heavy isotope enrichments that are built up in the cold phase.    When the star is born these organics increase in complexity and are revealed by evaporation in the hot (T $>$ 100~K) gas.  Based on spectral surveys, $\sim 1$\% of the carbon is found as organics in ices/vapor, depending on the temperature (Crockett et al. 2013, in prep.; Neill et al. 2013, in prep.).     When the young disk is born there is the potential for reprocessing as this is one of the least characterized stages.    The chemical and physical conditions of the forming disk are thus still highly uncertain and some aspects of the meteoritic record (e.g. calcium-aluminum rich inclusions) might be set in this stage.     This is a fruitful area of research in the coming years. 
 
 In the disk there is strong potential for both destruction and growth of organic molecules.   Solid state carbon
 compounds dominate the opacity from the near UV to the near infrared where most of the
stellar radiation originates \citep{dch01}.  Thus the  disk emission depends on the carbon abundance and distribution.    Models of disks with gaps or holes find that silicates exist closer the star than the carbonaceous materials which are present at larger radii \citep{Espaillat11}.  The emission from polycyclic aromatic hydrocarbons (PAHs)
 is also detected, but appears reduced in abundance when compared to the interstellar medium \citep{geers06}. 
 In the inner disk it is theorized that carbon-bearing grains, including PAHs, are oxidized and thus are not part of the terrestrial planet formation \citep{gail02, lbn10, Kress10}.  This is consistent with the overall depletion of carbon relative to silicon in CI chondrites compared to solar abundances \citep{Geiss87, lbn10}.  By inference this carbon must be in the gas.
 The abundance of CO has yet to be characterized and thus it is assumed to be a major carbon carrier inside of the CO condensation front.  
     At present we have no information on D/H ratios in ices.
  Beyond the detection of H$_2$CO, HCN, C$_2$H$_2$, and C$_3$H$_2$ the volatile organic content of disks is not known \citep{dutrey_ppv, cn08, salyk08, pascucci09, pont10, salyk11, qi13}. \\

{\it Oxygen}:  In the molecular cloud a large fraction of oxygen is locked in refractory silicates ($\sim 26$\%) and $\sim 31$\% in CO. 
Early in the evolution water ice and organics have not formed, thus prior to core condensation O is found either in neutral atomic form or in an unidentified refractory form \citep{whittet10}.    When the gas density exceeds $\gtrsim 10^4$~cm$^{-3}$ the water ice mantle forms, locking 20-40\% of this oxygen in the form of water ice.  When the star is born the water is provided to the disk as ice, which may remain as ice in the young disk and as vapor in the envelope \citep{Jorgensen10, Visser13}.    In the disk itself we have clear evidence for a transition from most of this water as vapor inside a few AU and then apparently to ice beyond \citep{zhang13}.   We have similar expectations for CO, which is another key oxygen carrier \citep{Qi13_sci}.   However, in the disk we do not have strong constraints on molecular abundances, beyond the general statement that in the colder regions of the disk molecular abundances appear lower than seen in the ISM, which is consistent with freeze-out onto grains \citep{dutrey_ppv}.
Furthermore, at present we have not identified all of the oxygen carriers throughout each stage; there may be an organic refractory component that includes significant oxygen \citep{whittet10}.\\

{\it Nitrogen}: Nitrogen is not a major constituent of dust grains and generally resides in the gas.   This should make tracking the main carriers easier.  However, one of the main potential carriers, N$_2$,  lacks a permanent dipole and cannot be probed by rotational transitions.
There are some observational constraints and general expectations.   In the molecular cloud phase, we know that the ionization potential of N is above 13.6 eV and major carriers require dust to shield from starlight, thus our expectation is that nitrogen remains in atomic form for general molecular cloud gas.   In the dense cores we observe gas phase NH$_3$, but it is not a major carrier.   Observations of N$_2$H$^+$, which forms from N$_2$ infer that in some instances most nitrogen is found in gas phase N$_2$ \citep{evd_ppiii}.   Detailed chemical modeling of Barnard 68 suggests instead that most nitrogen is found in neutral atomic form \citep{maret_n2}. \citet{Daranlot12} also argue that N$_2$ is not a major nitrogen reservoir, suggesting instead NH$_3$ in ice.   Towards  young protostars $\sim$10 - 20\% is inferred to be present as ammonia ice \citep{Bottinelli10, Oberg11c}, but the remaining form of N is uncertain and could be in atomic or molecular form.  For more complex molecules we do have estimates: HCN, and methyl/ethyl/vinyl cyanide are the main carriers of nitrogen comprising 14.6\% of available nitrogen (Crockett et al. 2013, in prep.).
The main form of nitrogen in the disk is unknown.   Since terrestrial worlds and even comets are highly nitrogen-poor \citep{wte91, Marty12} it is likely that nitrogen was in a very volatile form, perhaps N or N$_2$.  \citet{Marty12} suggests that nitrogen might have been delivered to the Earth in the form of organics.

\end{itemize}
\begin{figure}
\caption{Schematic summarizing the current state of knowledge regarding the development of chemical complexity throughout star and planet formation.  Image credits:  top left is the Barnard 68 pre-stellar core - ESO and \citet{alves_b68}; top middle is the L1527 protostellar and its disk (seen in silhouette) from \citet{Tobin12};  top right is an artist impression of a protoplanetary disk from ESO; Bottom left are solar system objects with image credit to NASA.
}
\includegraphics[height=0.5\textheight]{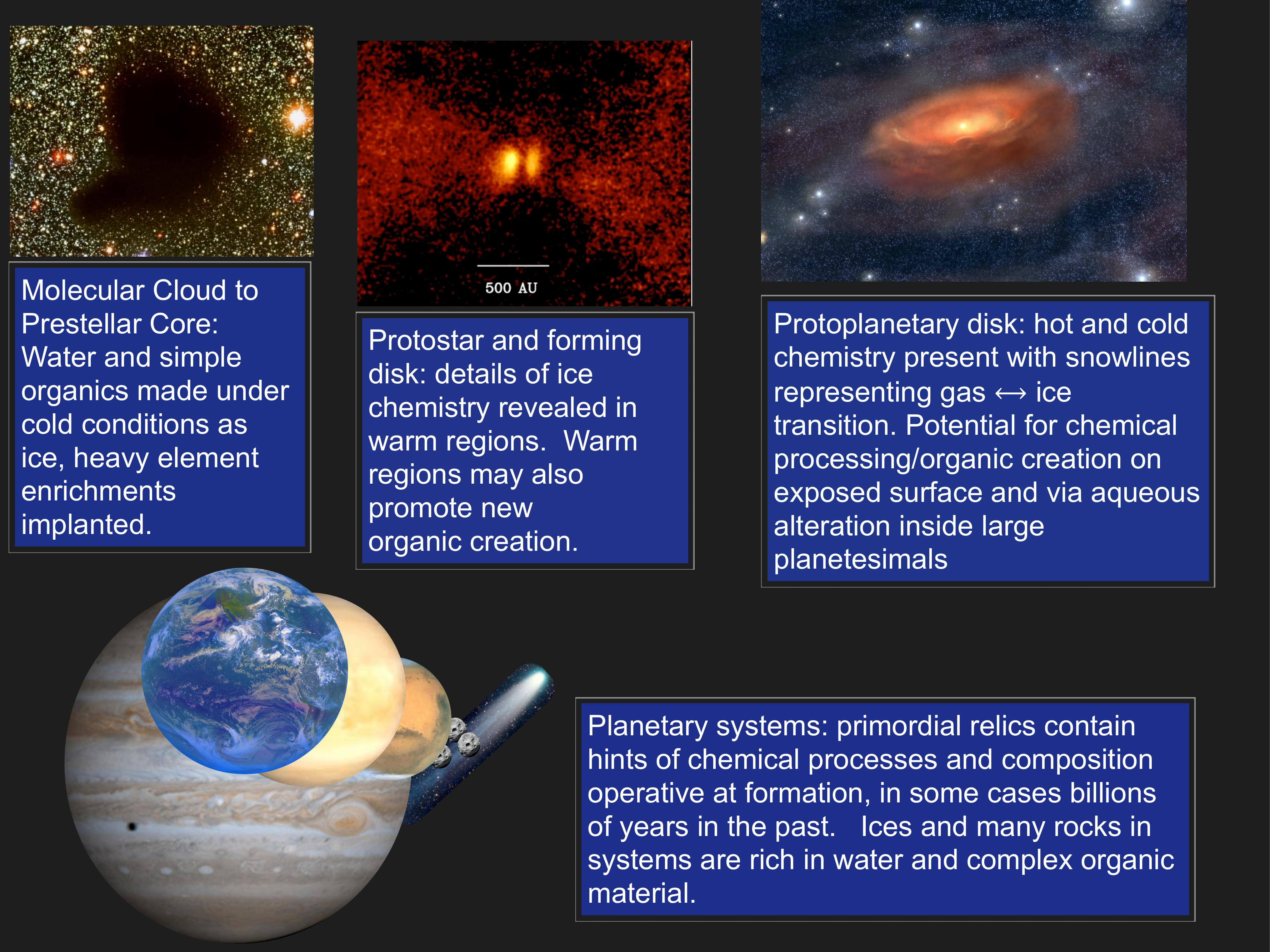}
\label{fig:summ}
\end{figure}

\section{Astrobiology:  A Marriage of Many Fields
}

Decades of transformative work has led us to a conversation about life's origin encompassing many fields: chemistry, biology, geology, physics, and astronomy.  From the astronomical perspective we have made significant progress in characterizing the beginning.  
Fig.~\ref{fig:summ} provides a graphical representation summarizing our current knowledge, which we detail below. 

  Prior to and during stellar birth we have information on the disposition within the carbon, hydrogen, oxygen, and nitrogen pools.  Water is formed before the star is born with high D/H ratios.   Thus the water that is supplied to the young planet potentially originated 500,000 years before the birth of the solar system.    During these phases the initial organics are born, also with the large deuterium enrichments. Some organics, e.g. PAHs, are inferred to be present, potentially in large abundance, but currently have no specific molecular assignment.   The organics that can be directly identified are fairly simple compared to, for example, the amino acids with measured heavy isotope enrichments in meteorites.  Organics are important carriers of carbon, but also nitrogen.  
  The chemistry that precedes and is associated with stellar birth thus readily produces simple monomers and copious amounts of water ice, with enrichments of heavy isotopes.
These are then provided by collapse to the young disk.  This represents the starting point for any subsequent processing within the disk and also deep inside meteorite parent bodies where reactions can proceed in some instances by aqueous alteration.    Perhaps just as importantly the physical properties seen towards the organic factories surrounding high mass stars (mass greater than several solar masses) are replicated somewhere in the young planet-forming disk.   Thus they serve as templates to understand chemical processes that will be active during the stage of planetary birth.    

Looking forward, the upcoming data from ALMA, and eventually the successor to the Hubble Space Telescope - the James Webb Space Telescope (JWST) - will provide data on the missing links in this process.    A number of important questions need to be addressed.  What processing occurs when the disk is born and in the disk itself?    We have fantastic theories, and a detailed record in the  solar system, but observational characterization of extra-solar systems is only just beginning.    Mixing of hot and cold materials appears to be a tale in our solar system; is this important for organic creation/alteration, changing the D/H ratios, and can any effects be identified?     We have now begun to potentially even detect young giant planets embedded in disks \citep{Kraus12, Quanz13}, and thus are truly studying planet formation in situ.   How to connect this to prebiotic chemistry somewhere on a young planet will remain a long-term challenge but remains an exciting goal.

\begin{theacknowledgments}
The author acknowledges support from NASA through the Origins of Solar Systems program (grant \# NNX12AI93G) and by the NSF via grant AST-1008800.   We are also grateful to Ilse Cleeves and Ruud Visser for a commenting on the manuscript.
\end{theacknowledgments}


\newcommand{\nat}{{ Nature }}
\newcommand{\aap}{{ Astron. \& Astrophys. }}
\newcommand{\aapr}{{ Astron. \& Astrophys. Rev.}}
\newcommand{\aj}{{ Astron.~J. }}
\newcommand{\apj}{{ Astrophys.~J. }}
\newcommand{\araa}{{ Ann. Rev. Astron. Astrophys. }}
\newcommand{\apjl}{{ Astrophys.~J.~Letters }}
\newcommand{\gca}{{ Geochim. Cosmochim. Acta}}
\newcommand{\ssr}{{ Space Science Rev.}}
\newcommand{\apjs}{{ Astrophys.~J.~Suppl. }}
\newcommand{\apss}{{ Astrophys.~Space~Sci. }}
\newcommand{\icarus}{{ Icarus }}
\newcommand{\mnras}{{ MNRAS }}
\newcommand{\pasp}{{ Pub. Astron. Soc. Pacific }}
\newcommand{\planss}{{ Plan. Space Sci. }}
\newcommand{\physrep}{{ Phys. Rep.}}
\newcommand{\bain}{{Bull.~Astron.~Inst.~Netherlands }}
\newcommand{\jgr}{{J. Geophys. Res.}}

\doingARLO[\bibliographystyle{aipproc}]
          {\ifthenelse{\equal{\AIPcitestyleselect}{num}}
             {\bibliographystyle{arlonum}}
             {\bibliographystyle{arlobib}}
          }
\bibliography{ted}

\end{document}